\documentclass[final,5p,times,twocolumn]{elsarticle}
\usepackage{natbib}
\usepackage{amssymb}
\usepackage{amsmath}
\usepackage{amsfonts}
\usepackage{color}
\usepackage{epstopdf}
\usepackage{hyphenat}

\newcommand{\nind}[1]{{(#1)}}

\begin{document}

\begin{frontmatter}
	
	
	\author{Eero Satuvuori\corref{cor1}\fnref{label1,label2,label3}}
	\ead{eero.satuvuori@unifi.it}
	\author{Mario Mulansky\corref{cor1}\fnref{label1}}
	\ead{mario.mulansky@isc.cnr.it}
	\author{Nebojsa Bozanic\corref{cor1}\fnref{label1}}
	\ead{nebojsa.bozanic@isc.cnr.it}
	\author{Irene Malvestio\corref{cor1}\fnref{label1,label2,label4}}
	\ead{irene.malvestio@upf.edu}
	\author{Fleur Zeldenrust\corref{cor1}\fnref{label5}}
	\ead{f.zeldenrust@neurophysiology.nl}
	\author{Kerstin Lenk\corref{cor1}\fnref{label6,label7}}
	\ead{lenk.kerstin@gmail.com}
	\author{Thomas Kreuz\corref{cor1}\fnref{label1}}
	\ead{thomas.kreuz@cnr.it}
	
	\fntext[label1]{Institute for Complex Systems, CNR, Sesto Fiorentino, Italy}
	\fntext[label2]{Department of Physics and Astronomy, University of Florence,  
		Sesto Fiorentino, Italy}
	\fntext[label3]{MOVE Research Institute, Department of Human Movement
		Sciences, Vrije Universiteit Amsterdam, The Netherlands}
	\fntext[label4]{Department of Information and Communication Technologies, Universitat Pompeu Fabra, Barcelona, Spain}
	\fntext[label5]{Donders Institute for Brain Cognition and Behavior, Radboud Universiteit, Nijmegen, The Netherlands}
	\fntext[label6]{BioMediTech, Tampere University of Technology, Tampere, Finland}
	\fntext[label7]{DFG-Center for Regenerative Therapies Dresden, Technische Universität Dresden, Dresden, Germany }
	

	\title{Measures of spike train synchrony for data with multiple time scales}
	
	
	
	\begin{abstract}

\textit{Background:} Measures of spike train synchrony are widely used in both experimental
and computational neuroscience.
Time-scale independent and parameter-free measures, such as the ISI-distance, the SPIKE-distance
and SPIKE-synchronization, are preferable to time scale parametric measures, since by
adapting to the local firing rate they take into account all the time scales of a given dataset. 

\noindent \textit{New Method:} In data containing multiple time scales (e.g. regular spiking
and bursts) one is typically less interested in the smallest time scales and a more adaptive
approach is needed.
Here we propose the A-ISI-distance, the A-SPIKE-distance and A-SPIKE-synchronization,
which generalize the original measures by considering the local relative to the global
time scales.
For the A-SPIKE-distance we also introduce a rate-independent extension called the
RIA-SPIKE-distance, which focuses specifically on spike timing.

\noindent \textit{Results:} The adaptive generalizations A-ISI-distance and A-SPIKE-distance
allow to disregard spike time differences that are not relevant on a more global scale.
A-SPIKE-synchronization does not any longer demand an unreasonably high accuracy for
spike doublets and coinciding bursts. Finally, the RIA-SPIKE-distance proves to be
independent of rate ratios between spike trains.

\noindent \textit{Comparison with Existing Methods:} We find that compared to the original
versions the A-ISI-distance and the A-SPIKE-distance yield improvements for spike trains
containing different time scales without exhibiting any unwanted side effects in other examples.
A-SPIKE-synchronization matches spikes more efficiently than SPIKE-Synchronization.

\noindent \textit{Conclusions:} With these proposals we have completed the picture, since we
now provide adaptive generalized measures that are sensitive to firing rate only (A-ISI-distance),
to timing only (ARI-SPIKE-distance), and to both at the same time (A-SPIKE-distance).
	\end{abstract}
	
	
\end{frontmatter}

\section{Introduction}

In neuroscience the neuronal action potential and its complex molecular
behavior \citep{Bear07} is often reduced to time-discrete events called
\textit{spikes}.
Due to the all-or-nothing paradigm of neurons together with the long silent
periods, the time stamps of the spike events are considered to be an accurate
enough description of the neuronal membrane potential \citep{QuianQuiroga13}.
These sequences of consecutive spikes are called \textit{spike trains}.
While spike trains do not directly provide information about the connections
between neurons, some form of link between two neurons is often inferred by
the similarity of their spike trains.
A spike train distance does not take into account the specific type of linkage,
but simply quantifies how (dis)similar the two spike trains are.
This makes spike train distances universal and as such they can be applied to all
systems that can be reduced to point processes.
\textcolor{black}{In addition to the obvious neuroscience applications, they have
already been used to study inter-personal coordination \citep{Rabinowitch15} and
social cognition \citep{Zapata16} among many other fields.}

Over the years many different measures have been developed in order to quantify
similarities between two or more spike trains \textcolor{black}{(see \cite{Victor15},
\cite{Naud11} and \cite{Kreuz11b} for an overview)}.
The two most known time scale parametric measures, the Victor-Purpura \citep{Victor96}
and the van Rossum distance \citep{VanRossum01}, describe spike train (dis)similarity
based on user-defined time scales to which the measures are mainly sensitive.
One drawback of these measures is the fixed time scale, since it sets a boundary
between rate and time coding for the whole recording.
However, for real data which typically contain many time scales (such as regular
spiking and bursts), this is difficult to detect with a measure that is mainly
sensitive to only one of them \citep{Chicharro11}.

The problem of having to choose one time scale has been eliminated in the three
time-resolved and time scale independent measures ISI\hyp{}distance \citep{Kreuz07c,Kreuz09},
SPIKE\hyp{}distance \citep{Kreuz11,Kreuz13} and SPIKE\hyp{}synchronization \citep{Kreuz15}.
The ISI\hyp{}distance \citep{Kreuz07c} is a measure of instantaneous rate dissimilarity.
It uses the interspike intervals (ISIs) to estimate the local firing rate of spike trains
and quantifies their differences in a time-resolved manner.
The SPIKE\hyp{}distance \citep{Kreuz11} compares the spike time accuracy between
spike trains and uses instantaneous firing rates to adapt to the local time scale.
Finally, SPIKE\hyp{}synchronization \citep{Kreuz15} is a discrete time-resolved measure
of similarity based on ISI-derived coincidence windows that are used to determine
if two spikes from different spike trains are coincident or not. 
\textcolor{black}{These measures have already been successfully applied in many different contexts;
for example they have been used to detect determinism in point processes \citep{Andrzejak14},
to find correlations between spike trains and behaviour in an inverse neurocontroller
\citep{Dura-Bernal16} and to evaluate a bio-inspired locomotion system in robotics \citep{Espinal16}.}

Since they always adapt to the local firing rate, all three of these measures are
time scale free.
While they correctly identify the relative firing rate differences, they have no
concept of actual time scales and treat all time scales as equally important.
This has the consequence that for very small time scales even minor deviations from
perfect synchrony lead to very high values of dissimilarity.
However, for real data the smallest time scales are often not very relevant and any
dissimilarities there can mostly be disregarded.
Thus in this case the measures' focus on the local time scales results in a (spurious)
amplification of dissimilarities which compared to the global time scales are rather
negligible.

Here we address this problem by proposing generalizations to the three measures
called adaptive ISI\hyp{}distance (A-ISI-distance), adaptive SPIKE\hyp{}distance
(A-SPIKE-distance) and adaptive SPIKE\hyp{}synchronization (A-SPIKE-synchronization).
These generalized definitions add a notion of the relative importance of local differences 
compared to the global time scales.
In particular, they start to gradually ignore differences between spike trains for ISIs
that are smaller than a minimum relevant time scale (MRTS).
The MRTS is implemented by an additional variable $\mathcal{T}$ which can either be defined as
a parameter or estimated directly from the data.

In some neuroscience applications only the similarity of spike timing is important and
rate differentiation is not a desired property.
While the A-ISI-distance is sensitive to firing rate alone and the A-SPIKE-distance
responds to differences in both rate and timing, there is currently no measure that
focuses only on spike timing.   
Therefore, in a second step we extend the A-SPIKE-distance into the rate-independent adaptive
SPIKE\hyp{}distance (RIA-SPIKE-distance) which still identifies spike time differences but ignores
any rate deviations between the spike trains.

The remainder of this paper is organized as follows.
In Section \ref{s:Adaptive} we describe the generalized definitions of the three measures,
the A-ISI-distance (Section \ref{ss:ISI-dist}), the A-SPIKE-distance (Section \ref{ss:SPIKE-dist}),
and A-SPIKE-synchronization (Section \ref{ss:SPIKE-sync}).
In Section \ref{ss:MRTS_thr} we introduce a way to estimate the threshold value directly
from the data.
We then investigate \textcolor{black}{using both simulated and real data} how both
the original measures and the adaptive generalizations deal with multiple time scales
(Section \ref{ss:Results}).
In Section \ref{s:RI} we add a rate-independent extension to A-SPIKE-distance (Section
\ref{ss:RIA-SPIKE}) and afterwards study the effects of the extension (Section
\ref{ss:RIResults}).
The implications of the extensions are discussed in Section \ref{s:Discussion}.
Finally, in the \ref{Appendix-Methods} we cover some non-trivial subtleties of
the definitions for all three measures.
First we provide the definitions for the periods before the first and after the 
last spike in a spike train (where the interspike interval is not defined), 
and then we deal with the two special cases of empty spike trains and spike trains
with only one spike.
The \textcolor{black}{two experimental datasets} used in Section \ref{ss:Results} are
described in \ref{Appendix-Data}.

\section{\label{s:Adaptive}Adaptive Generalizations}

In this Section we introduce the adaptive generalizations of the established measures
ISI\hyp{}distance \citep{Kreuz07c}, SPIKE\hyp{}distance \citep{Kreuz11} and
SPIKE\hyp{}synchronization \citep{Kreuz15}, which we will call A-ISI-distance,
A-SPIKE-distance, and A-SPIKE-synchronization.
All three generalizations are built on a minimum relevant time scale (MRTS) which is
implemented via the threshold parameter $\mathcal{T}$.
This threshold is used to determine if a difference between the spike trains should
be assessed in a local context or in relation to the global time scales.
This threshold is used for all three measures, but the way it is applied varies.
The generalized measures fall back on the original definitions when $\mathcal{T} = 0$.
In the following this is what we refer to whenever we talk of the original measures.
In this case even the smallest time scales matter and all differences are assessed
in relation to the local context only.

Note that the upcoming definitions only apply to the interval between the first
and the last spike.
In \ref{ss:ISI_SPIKE_edge_corrections} and \ref{ss:SYNC_edge_corrections} they will be
completed to range from the start of the recording $t_s$ to the end of the recording $t_e$ .
Equally, some of the following equations are ill-defined when there are less than
two spikes in a spike train.
These special cases will be handled in \ref{ss:ISI_SPIKE_special} and \ref{ss:SYNC_special}.

Throughout the paper we denote the number of spike trains by $N$, indices of
spike trains by $n$ and $m$, spike indices by $i$ and $j$ and the number of spikes
in spike train $n$ by $M_n$.
The spike times of spike train $n$ are denoted by $\{t^{(n)}_i\}$ with $i=1,\dots,M_n$.

\begin{figure}[t]
	\includegraphics[width = 0.48\textwidth]{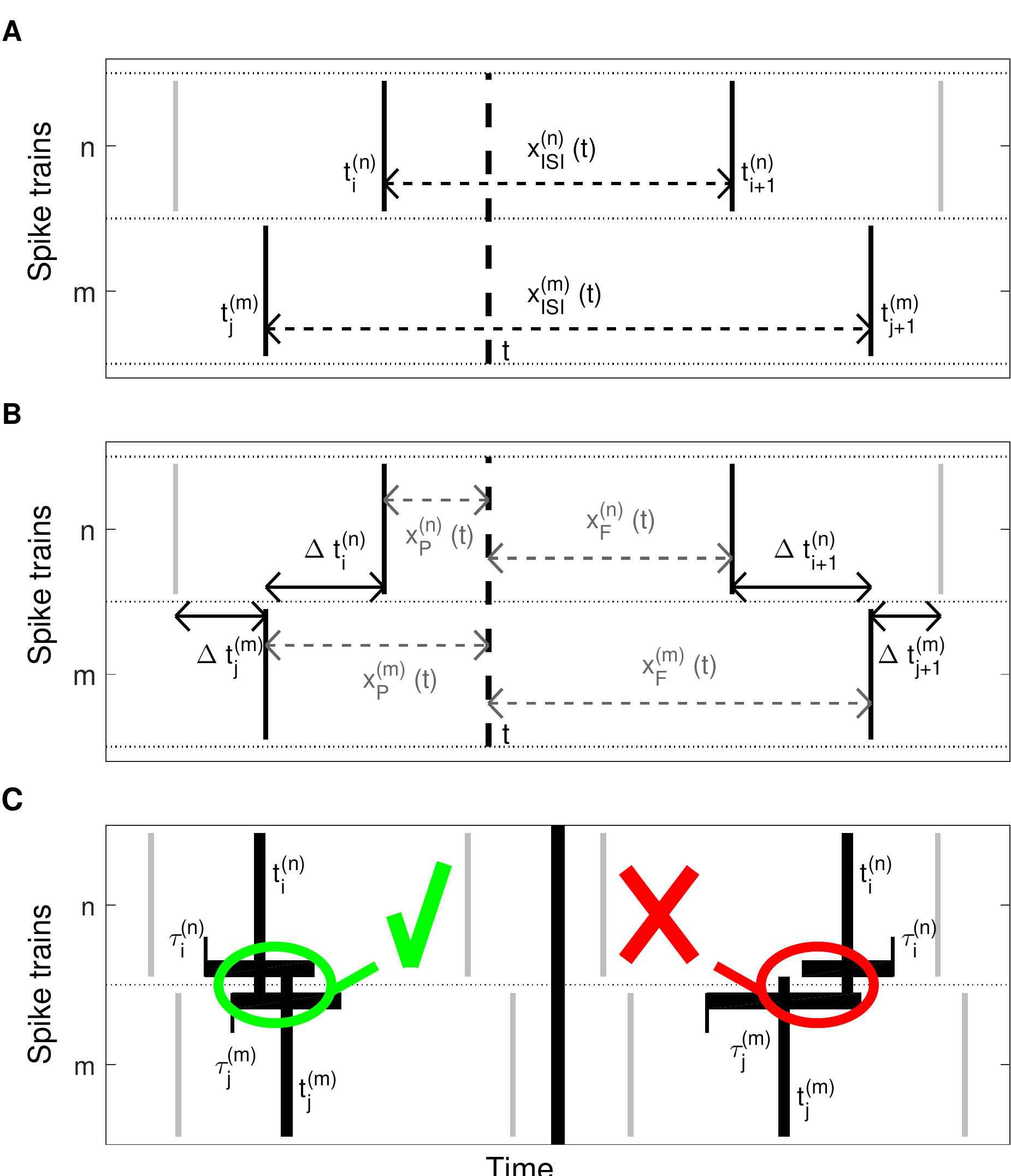}
	\caption[Figure]{
		Schematic drawing for all three measures. 
		(A) Illustration of the variables that define the ISI\hyp{}distance.
		The instantaneous interspike intervals $x_{ISI}^{(n)}(t)$ are used as estimates of
		the local firing rate.
		(B) Additional variables employed in the definition of the SPIKE\hyp{}distance.
		(C) Coincidence criterion for SPIKE\hyp{}synchronization.
		The coincidence window of each spike is derived from its two surrounding interspike
		intervals.
		Here we illustrate two different examples.
		The two spikes on the left side are considered coincident since both lie in each
		other's coincidence windows.
		On the right there is no coincidence since the spike from the second spike train
		is outside of the coincidence window from the spike of the first spike train.
		\label{Fig:Illustration}}
\end{figure}

\subsection{\label{ss:ISI-dist}Adaptive ISI-distance}

The A-ISI-distance measures the instantaneous rate difference between spike trains (see Fig.
\ref{Fig:Illustration}A).
It relies on a time-resolved profile, meaning that a dissimilarity value is defined for
each time instant.
To obtain the profile, we assign to each time instant $t$ the time of the previous spike
\begin{equation} \label{eq:Prev-Spike}
	t_{\mathrm {P}}^{(n)} (t) = \max\{t_i^{(n)} | t_i^{(n)} \leq t\}  \quad
	\textrm{for }\quad t_1^{(n)} \leqslant t \leqslant t_{M_n}^{(n)}
\end{equation}
and the time of the following spike
\begin{equation} \label{eq:Foll-Spike}
	t_{\mathrm {F}}^{(n)} (t) = \min\{t_i^{(n)} | t_i^{(n)} > t\} \quad \textrm{for } \quad
	t_1^{(n)} \leqslant t \leqslant t_{M_n}^{(n)}.
\end{equation}
From this for each spike train $n$ an instantaneous ISI can be calculated as
\begin{equation} \label{eq:ISI}
	x_{\mathrm{ISI}}^{(n)} (t) = t_{\mathrm {F}}^{(n)} (t) - t_{\mathrm {P}}^{(n)} (t).
\end{equation}

For the A-ISI-distance we define the MRTS such that when the ISIs of both spike trains
are smaller than a threshold value $\mathcal{T}$, this value is used instead.
The pairwise A-ISI-profile is then defined as

\begin{equation} \label{eq:pairwise A-ISI-profile}
I_{n,m}^A (t) = \frac{|x_{\mathrm{ISI}}^{(n)} (t) - x_{\mathrm{ISI}}^{(m)}
	(t)|}{\max \{x_{\mathrm{ISI}}^{(n)} (t), x_{\mathrm{ISI}}^{(m)} (t),\mathcal{T}\} }.
\end{equation}
The multivariate A-ISI-profile is obtained by averaging over all pairwise A-ISI-profiles
\begin{equation} \label{eq:multivariate ISI}
	I^A (t) = \frac{2}{N(N-1)}\sum_{n=1}^{N-1} \sum_{m=n+1}^N I_{n,m}^A (t).
\end{equation}
This is a non-continuous piecewise constant profile and a final integration over time
gives the A-ISI-distance
\begin{equation} \label{eq:ISI-distance}
	D_I^A = \frac{1}{t_e-t_s} \int_{t_s}^{t_e} I^A (t)dt .
\end{equation}
If the threshold $\mathcal{T}$ is set to zero, the generalized ISI\hyp{}distance $D_I^A$ falls back
to the original ISI\hyp{}distance $D_I$.

\subsection{\label{ss:SPIKE-dist}Adaptive SPIKE-distance}

The A-SPIKE-distance measures the accuracy of spike times between spike trains relative to
local firing rates (see Fig. \ref{Fig:Illustration}B).
In order to assess the accuracy of spike events, each spike is assigned the distance to its
nearest neighbor in the other spike train
\begin{equation} \label{eq:Closest spike}
	\Delta t_i^{(n)} = \min_j(|t_i^{(n)} - t_j^{(m)}|).
\end{equation}
These distances are then interpolated between spikes using for all times $t$ the time differences
to the previous spike
\begin{equation} \label{eq:Weighting_prev}
	x_P^{(n)}(t) = t-t_i^{(n)}    \quad \textrm{for }\quad  t_i^{(n)} \leqslant t \leqslant
	t_{i+1}^{(n)},
\end{equation}
and to the following spike
\begin{equation} \label{eq:Weightings_foll}
	x_F^{(n)}(t) = t_{i+1}^{(n)}-t \quad \textrm{for }\quad  t_i^{(n)} \leqslant t \leqslant
	t_{i+1}^{(n)}.
\end{equation}
These two quantities define a time-resolved dissimilarity profile from discrete values the
same way as Eqs. \ref{eq:Prev-Spike} and \ref{eq:Foll-Spike} did for the A-ISI-distance.
The instantaneous weighted spike time difference for a spike train can then be calculated
as the interpolation from one difference to the next
\begin{equation} \label{eq:weighted distance}
	S_{n} (t) = \frac{\Delta t_i^{(n)} (t)x_F^{(n)}(t) + \Delta t_{i+1}^{(n)}
		(t)x_P^{(n)}(t)}{x_{\mathrm{ISI}}^{(n)} (t)}\quad,\quad  t_i^{(n)} \leqslant t \leqslant
	t_{i+1}^{(n)}.
\end{equation}
This function is analogous to the term ${x_{\mathrm{ISI}}^{(n)}}$ for the ISI\hyp{}distance,
with the only difference that it is piecewise linear instead of piecewise constant.
It is also continuous.

The pairwise A-SPIKE-distance profile is obtained by averaging the weighted spike time differences,
normalizing to the local firing rate average and, finally, weighting each profile by the
instantaneous firing rates of the two spike trains
\begin{equation} \label{eq:A-SPIKE profile}
S_{m,n}^{A} (t) = \frac{S_n x_{\mathrm{ISI}}^m (t) +S_m x_{\mathrm{ISI}}^n (t) }{2\langle
	x_{\mathrm{ISI}}^{n,m} (t)\rangle \max\{ \langle x_{\mathrm{ISI}}^{n,m} (t)\rangle, \mathcal{T} \} }.
\end{equation}
We define the MRTS by using a threshold, that replaces the denominator of weighting to spike time
differences if the mean is smaller than the threshold $\mathcal{T}$. 
This profile is analogous to the pairwise A-ISI-profile ${I_{n,m}^A (t)}$, but again it is 
piecewise linear, not piecewise constant.
Unlike $S_{n} (t)$ it is not continuous, since typically it exhibits instantaneous jumps at the
times of the spikes.
The multivariate A-SPIKE-profile is obtained the same way as the multivariate A-ISI-profile,
by averaging over all pairwise profiles
\begin{equation} \label{eq:SPIKE profile}
	S^A (t) = \frac{2}{N(N-1)}\sum_{n=1}^{N-1} \sum_{m=n+1}^N S_{m,n}^{A} (t).
\end{equation}
Finally, also the A-SPIKE-distance is calculated as the time integral over the multivariate
profile
\begin{equation} \label{eq:Spike-Distances}
	D_S^A = \frac{1}{t_e-t_s} \int_{t_s}^{t_e} S^A(t)dt.
\end{equation}
For $\mathcal{T} = 0$ also the A-SPIKE-distance falls back to the SPIKE\hyp{}distance.

\begin{figure}[bt]
	\includegraphics[width = 0.48\textwidth]{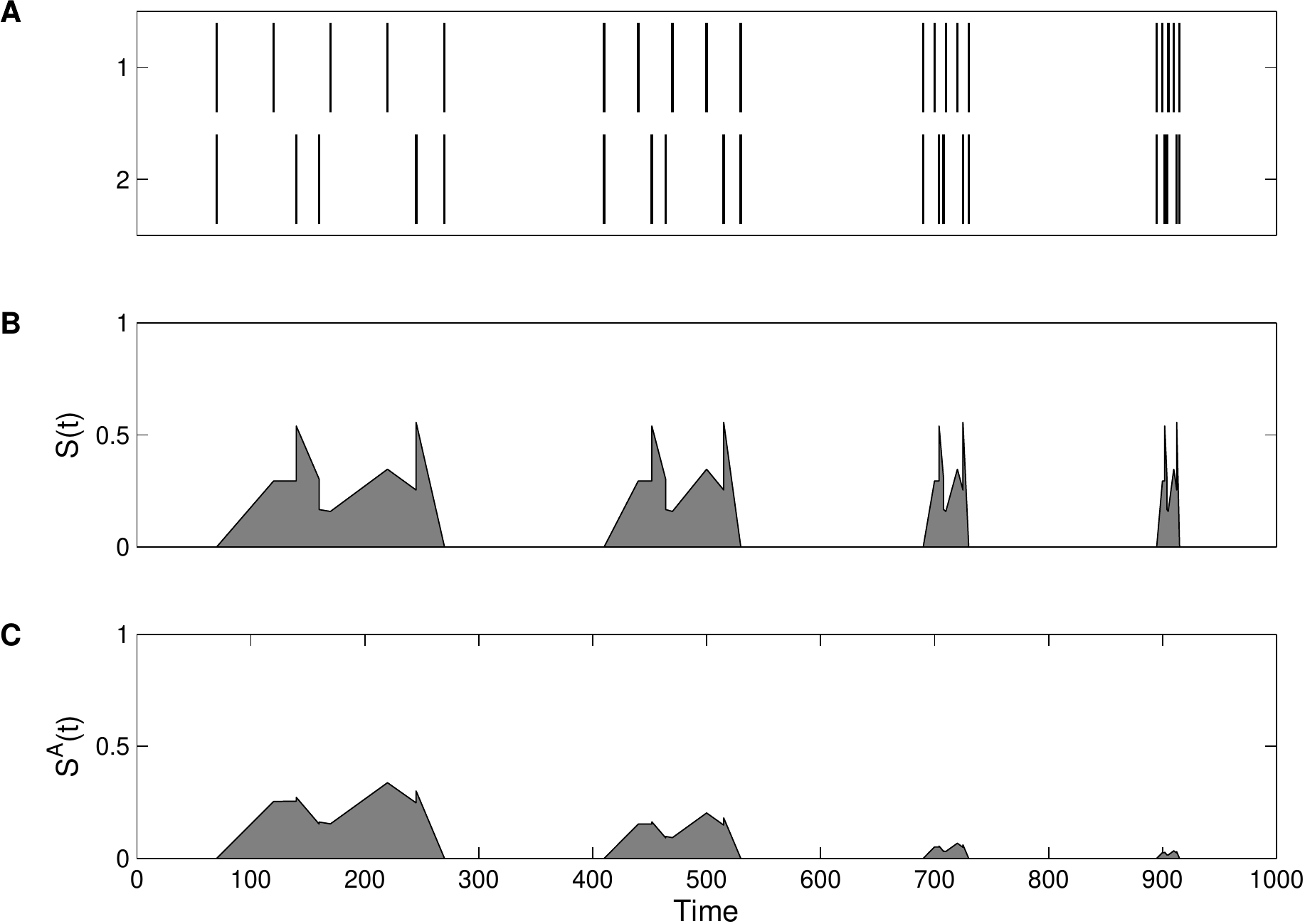}
	\caption[Figure]{
		An example spike train pair and its SPIKE\hyp{}distance and A-SPIKE-distance
		profiles.
		(A) Two spike trains consisting of four events with five spikes each.
		The sequence is the same for all four events, only the time scale is getting
		shorter and shorter.
		From a global perspective the first event consists of non-synchronous
		individual spikes, while the last event consists of coincident bursts.
		The two events in the middle are intermediates.
		(B) The SPIKE\hyp{}distance considers only the local context and thus the profile
		shape is the same for all four events.
		(C) The A-SPIKE-distance takes into account also the global time scales.
		Like the SPIKE\hyp{}distance it judges the first event as very dissimilar, but in
		contrast to the the SPIKE\hyp{}distance it scales down the small spike time differences
		in the bursts and thus considers the coincident burst in the last event as very
		similar.
		\label{motivation3}}	
\end{figure}

The effect of applying the threshold can be seen in Fig. \ref{motivation3}.
The first event of five spikes is compressed more and more until it becomes a single
burst in the fourth event. 
The original SPIKE\hyp{}distance profile $S(T)$ has the same proportions of dissimilarity
for all events, since it uses local context only and thus considers all time scales as
equal, while the A-SPIKE-distance profile $S^A(t)$ is scaled down when the differences
become small compared to the global time scales.

\subsection{\label{ss:SPIKE-sync}Adaptive SPIKE-synchronization}

A-SPIKE-synchronization quantifies how many of the possible spike coincidences in a dataset are
actually occurring (Fig. \ref{Fig:Illustration}C).
While the A-ISI-distance and the A-SPIKE-distance are measures of dissimilarity which
obtain low values for similar spike trains, A-SPIKE-synchronization measures similarity.
If all the spikes are coincident with a spike in all the other spike trains, its
value will be one.
In contrast, if none of the spikes are coincident, it will be zero.

The original SPIKE\hyp{}synchronization \citep{Kreuz15} is parameter- and time scale-free, since it
uses the adaptive coincidence detection first proposed for the measure event synchronization
\citep{QuianQuiroga02b}. 
The coincidence window, i.e., the time lag below which two spikes from two different
spike trains, $t_i^{(n)}$ and $t_j^{(m)}$, are considered to be coincident, is adapted
to the local firing rate.
Spikes are coincident only if they both lie in each other's coincidence windows.

For A-SPIKE-synchronization we generalize the definition by introducing a threshold, which
decides if the window is determined locally or if the global time scales should be taken into
account.
As a first step, we define the ISI before the spike as
\begin{equation} \label{eq:ISI_before}
	x_{iP}^{(n)}= \lim_{t \to t_i-} x_{\mathrm{ISI}}^{(n)} (t)
\end{equation}
and the ISI after the spike as
\begin{equation} \label{eq:ISI_after}
	x_{iF}^{(n)}= \lim_{t \to t_i+} x_{\mathrm{ISI}}^{(n)} (t).
\end{equation}
The coincidence window for spike $i$ of spike train $n$ is defined by determining the
minimum coincidence window size for a spike as half the length of the two ISIs adjacent
to the spike
\begin{equation}\label{eq:A-councidence}
	\tau_{i}^{(n)} = \frac{1}{2}\min \{ x_{iP}^{(n)},	x_{iF}^{(n)} \},
\end{equation}
and allowing asymmetric coincidence windows based on MRTS.
This is done by replacing $\tau_{i}^{(n)}$ with the threshold value $\mathcal{T}$,
if it is the smaller of the two. 
Since the threshold value is derived from ISIs and the coincidence window spans both
sides of the spike, only half of the threshold spans each side.
For the A-ISI- and the A-SPIKE-distance the changes induced by the threshold appear
gradually, but for A-SPIKE-synchronization they occur as an abrupt jump from 0 to 1.
Therefore, to compensate for the binary nature of A-SPIKE-synchronization, the
threshold is divided by two, resulting in an overall factor of $1/4$.
The coincidence windows of neighboring spikes are not allowed to overlap, and
thus each side is limited to half the ISI even if the threshold is larger.
Thus, the coincidence window before the spike is determined as
\begin{equation} \label{eq:A-CoincidenceP}
	\tau_{iP}^{(n)} = \min \{ \max (\frac{1}{4}\mathcal{T},
	\tau_{i}^{(n)}),\frac{1}{2}x_{iP}^{(n)} \}
\end{equation}
and the coincidence window after the spike as
\begin{equation} \label{eq:A-CoincidenceF}
	\tau_{iF}^{(n)} = \min \{ \max (\frac{1}{4}\mathcal{T},
	\tau_{i}^{(n)}),\frac{1}{2}x_{iF}^{(n)} \}.
\end{equation}
The combined coincidence window for spikes $i$ and $j$ is then defined as
\begin{equation} \label{eq:A-Coincidence-MaxDist}
\tau_{ij}^{(n,m)} = \begin{cases}
\min \{\tau_{iF}^{(n)},\tau_{jP}^{(m)}\}    & {\rm if} ~~ t_i \leqslant t_j \cr
\min \{\tau_{iP}^{(n)},\tau_{jF}^{(m)}\}      &  {\rm otherwise}
\end{cases}.
\end{equation}
\begin{figure}[tb]
	\includegraphics[width = 0.48\textwidth]{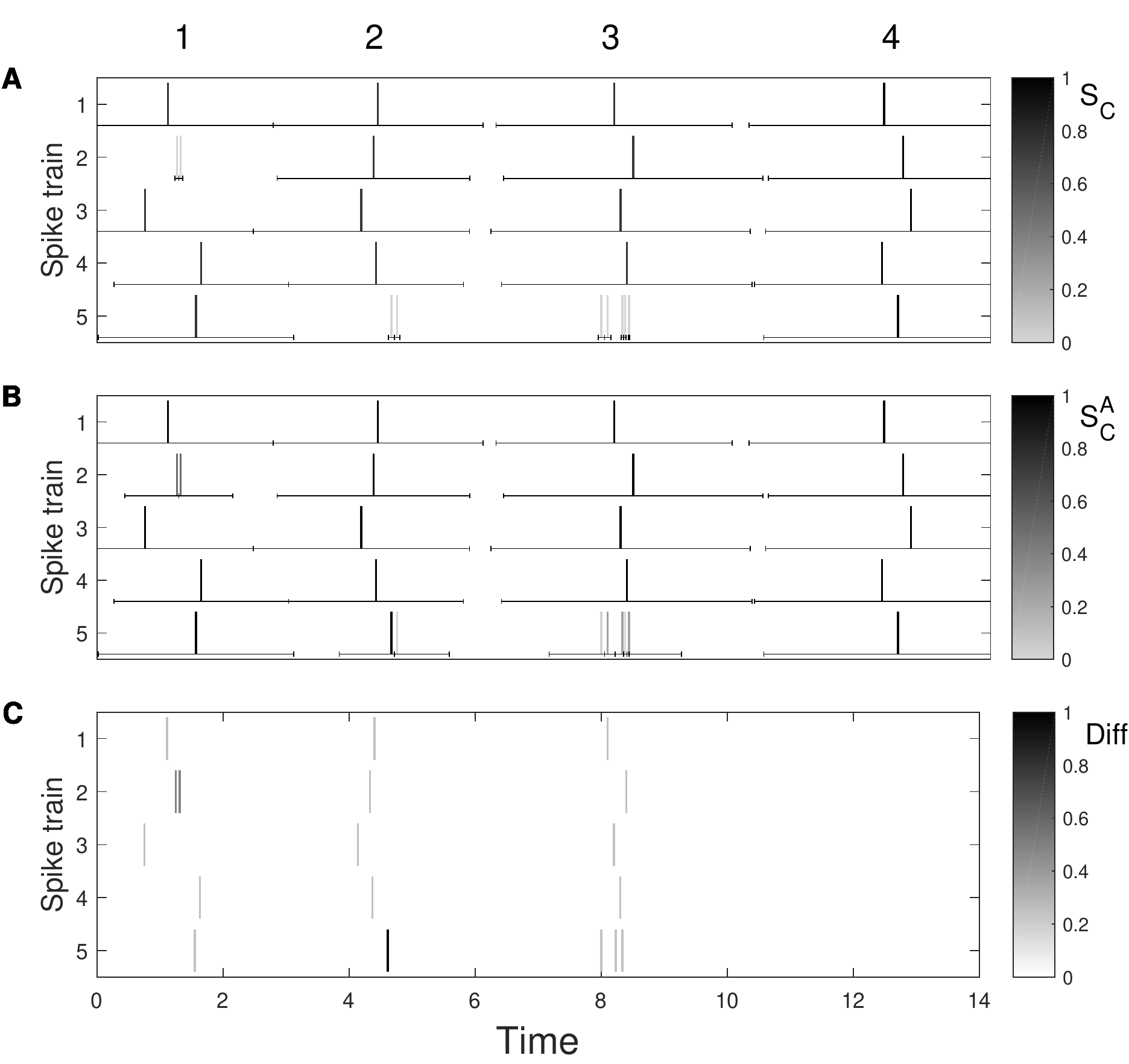}
	\caption{
		SPIKE\hyp{}synchronization (A), A-SPIKE-synchronization (B) and their difference (C)
		illustrated using five spike trains with four simple events. 		
		For the original measure (A) the small interspike intervals of spike doublets
		(first and second event) or bursts (third event) result in an unreasonably high
		demand for spike timing accuracy.
		With the adaptive generalization (B) for all these cases the likelihood increases
		that at least one of the spikes is part of a coincidence.
		On the other hand, if there are no doublets or bursts (last event), nothing changes
		(best seen in C).
		Note that the color scales differ, for better visibility we use grey-black in A
		and B but white-black in C. 
		\label{fig:A-SYNC_simple}}
\end{figure}

The coincidence criterion can be quantified by means of a coincidence indicator
\begin{equation} \label{eq:SPIKE-Coincidence}
	C_i^{(n,m)} = \begin{cases}
		1     & {\rm if} ~~ \min_j \{|t_i^{(n)} - t_j^{(m)}|\} < \tau_{ij}^{(n,m)} \cr
		0     & {\rm otherwise}
	\end{cases}.
\end{equation}
This definition ensures that each spike can only be coincident with at most one
spike in the other spike train.
The coincidence criterion assigns either a one or a zero to each spike depending
on whether it is part of a coincidence or not.
For each spike of every spike train, a normalized coincidence counter
\begin{equation} \label{eq:Multi-SPIKE-Coincidence}
	C_i^{(n)} = \frac1{N-1}\sum_{m\neq n} C_i^{(n,m)}
\end{equation}
is obtained by averaging over all $N-1$ bivariate coincidence indicators involving
the spike $i$ in spike train $n$.

This way we have defined a coincidence indicator for each individual spike in the
spike trains.
In order to obtain one combined similarity profile, we pool the spikes of the
spike trains as well as their coincidence indicators by introducing one overall
spike index $k$.
This yields one pooled set of coincidence indicators
\begin{equation} \label{eq:Multi-Profile}
	\{C_k\} = \bigcup_n \{C_i^{(n)} \}
\end{equation}
from which the A-SPIKE-synchronization profile $C^A (t_k)$ can be obtained via
$C^A (t_k) = C (k)$.
Finally, A-SPIKE-synchronization is defined as the average value of this discrete profile
\begin{equation} \label{eq:Multi-SPIKE-Synchronization}
	S_C^{A} = \frac{1}{M} \sum_{k=1}^M C^A (t_k),
\end{equation}
where $M$ is the overall number of spikes. 
In Fig. \ref{fig:A-SYNC_simple} we illustrate how the asymmetric coincidence
windows of A-SPIKE-synchronization allow for a better coverage of burst events
which makes it easier to match spikes when compared to the original SPIKE\hyp{}synchronization
(A-SPIKE-synchronization with $\mathcal{T} = 0$). 
It is important to note that reducing differences below threshold adds coincidences and thus,
since it is a measure of similarity, A-SPIKE-synchronization can only increase.

\subsection{\label{ss:MRTS_thr} Selecting the threshold value}

In neuroscience typical time scales are in the range of milliseconds or sometimes seconds
and any time scales below this will not be considered relevant.
In fields such as meteorology the respective time scales could be hours and days or even
months and years.
The relevant time scales clearly depend on the system under consideration.
Setting the minimum relevant time scale (MRTS) for a given dataset might not be a simple task.
To address this, we propose a method to extract a threshold value from the spike trains,
that is based on the proportions of the different time scales present in the data.

It is important to note that the selected MRTS is not an indicator of a
time scale of the system; it just determines the outcome of the adaptive
generalizations.
It is also not a hard set limiter neglecting everything below the threshold, but
rather it marks the time scale from which on differences are considered in the
global instead of the local context.
Thus from this time scale on deviations from synchrony are treated as less and less
relevant the smaller they get, even if they are large in relation to the local time scales.

\begin{figure}[tb]	
	\includegraphics[width = 0.48\textwidth]{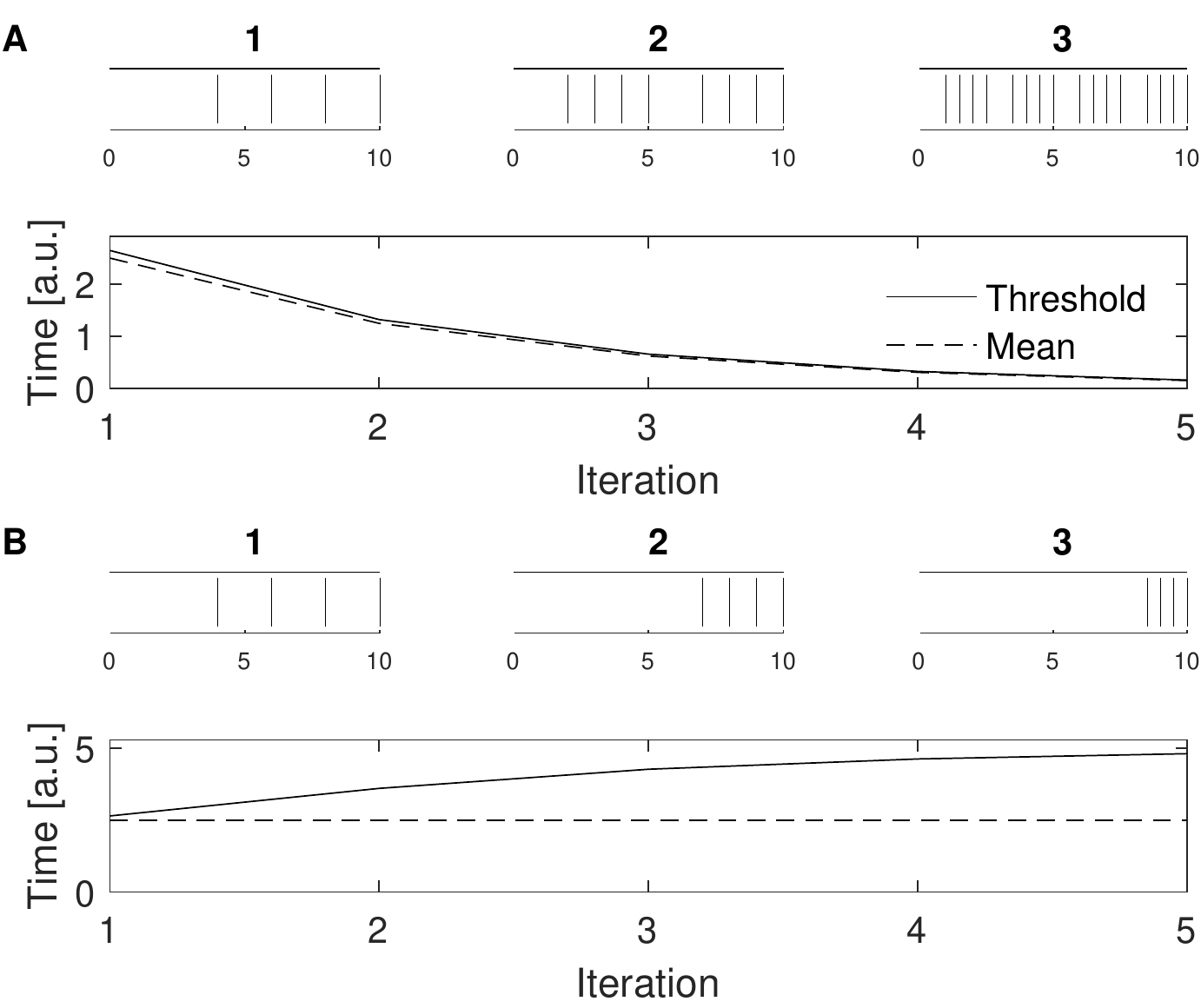}
	\caption[Figure]{Threshold value vs. the mean of the ISI-distribution.
		(A) Dependence on the number of spikes (first criterion).
		In each iteration the number of spikes is increased by concatenating two
		half-length copies of the previous iteration.
		Both the mean and the threshold decrease with spike count.
		(B) Dependence on the ISI-distribution (second criterion).
		From iteration to iteration the ISI-distribution is changed by halving the three
		short ISIs and prolonging the long ISI accordingly.
		Since the spike count (and thus the number of ISIs) is kept constant, the mean does
		not respond to this change.
		However, the threshold correctly increases with the heightened importance of the
		long ISI.
		\label{Fig:THR_figure}}
\end{figure}

The purpose of the threshold is to act as an indicator of what globally is a high rate or
inversely a small ISI.
The original normalizations are based on the ISIs, so it is reasonable to determine
the threshold from the pooled ISI-distribution.
We use the ISIs after the edge effect has been corrected (see \ref{ss:ISI_SPIKE_edge_corrections}).
The threshold should fulfill two main criteria.
First, it needs to decrease proportionally to the spike count, so that increasing
rates (or longer recordings with the same rate) do not change the threshold.
Second, the threshold should respond to changes in the ISI-distribution so that it is
able to adapt between single and multiple time scale data sets.
In Fig. \ref{Fig:THR_figure} we use a simple spike train motive of just four spikes
to illustrate these two criteria.

The most straightforward threshold would be the mean length of the ISIs
\begin{equation} \label{eq:mean_ISI}
	\langle {L_{\mathrm{ISI}}}\rangle = \frac{\sum_{g=1}^G L_{\mathrm{ISI}}^g
	}{M_{ISI}} = \frac{L}{M_{ISI}}.
\end{equation}
Here $L^g_{\mathrm{ISI}}$ denotes the ISI-length and $M_{ISI}$ is the total number of
ISIs in the pooled ISI-distribution.
In the numerator the sum of the lengths of all ISI equals the overall length $L$ of
the pooled ISIs.
Apart from edge effect corrections this is equal to the product of recording length
and number of spike trains which is a constant.
Thus while the mean of ISIs depends on the number of spikes (Fig. \ref{Fig:THR_figure}A),
for a given number of spikes (number of ISIs) it is completely independent of how the ISIs
are distributed around the mean (Fig. \ref{Fig:THR_figure}B).
It adapts to the spike count but not to the proportions in which the ISIs appear in
the data thus fulfilling the first but not the second criterion.

To fulfill both criteria one needs to not just count the interspike intervals but
weight them by their length.
This reduces the importance of short ISIs and allows the long ISIs to influence the
threshold according to their contribution and not just number.
It is equivalent to taking the mean of the second moments of the ISIs
\begin{equation}
\label{eq:mean_second_moment_ISI}
	\mathcal{T} = \sqrt{\langle {(L_{\mathrm{ISI}})^2}\rangle} =
	    \sqrt{\frac{\sum_{g=1}^G {L_{\mathrm{ISI}}^g}^2}{M_{ISI}}}.
\end{equation}
Note that in order to obtain a value with the right dimension the square root of
the average must be taken.
This threshold value has roughly the same dependence on the number of spikes as the mean
value (Fig. \ref{Fig:THR_figure}A), however, in contrast to the mean it is also
sensitive to changes in the ISI-distribution.
In summary, using $\mathcal{T}$ as the MRTS fulfills both criteria set for the threshold.

\subsection{\label{ss:Results}Results}

In this Section we investigate how both the adaptive generalizations (with automated
thresholding) and the original measures deal with multiple time scales.
For the A-ISI-distance and the A-SPIKE-distance we use a test spike train set consisting
of simulated and real spike trains to study the effect of the generalized versions
(Section \ref{ss:Results-A-ISI-SPIKE}).
After that, in Section \ref{ss:Results-A-SPIKE-sync}, we study on real MEA recordings
how A-SPIKE-synchronization differs from SPIKE\hyp{}synchronization.
\textcolor{black}{In Section \ref{ss:Results-Parameter-test} we systematically test the influence
of the amount of bursts on the difference between adaptive and original measures.
Finally, in Section \ref{ss:Results-Real_Data} we investigate how the adaptive versions change
the analysis of neuronal reliability in an experimental dataset.}

\subsubsection{\label{ss:Results-A-ISI-SPIKE}Adaptive ISI-distance and adaptive SPIKE-distance}

We address two points.
First, we look at the sensitivity of the adaptive generalizations and verify that
in the presence of bursts they perform better than the original measures.
Second, we also make sure that the changes are specific, e.g., we confirm
that in all other cases and especially if there are no bursts, the adaptive
generalizations do not exhibit unwanted side effects.

\begin{figure}[tb]
	\includegraphics[width = 0.48\textwidth]{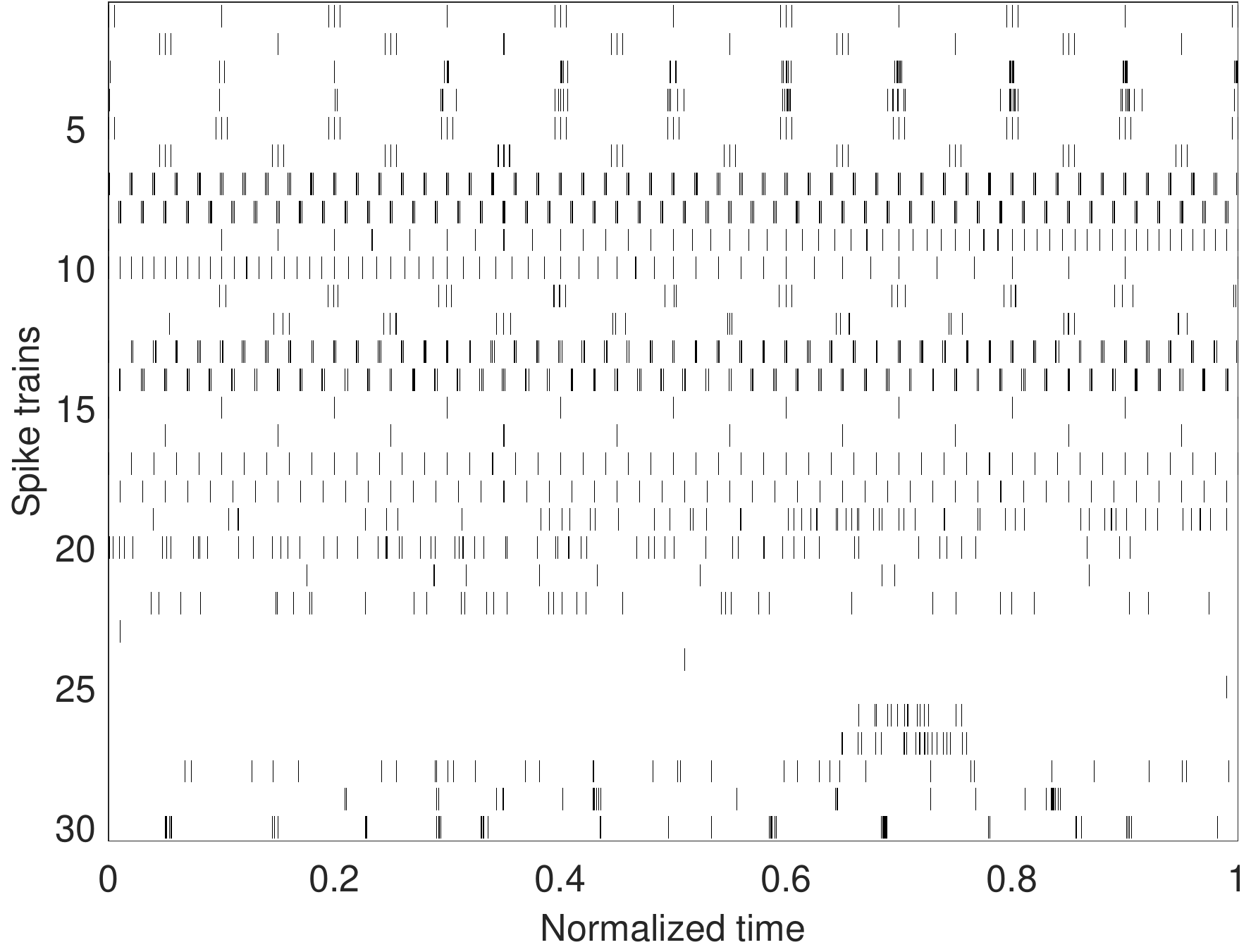}
	\caption[Figure]{
		Spike train test set used to compare the generalized versus the original measures
		of spike	 train synchrony.
		Spike trains 1-25 are artificially constructed examples which cover a range of
		archetypical spiking patterns, whereas spike trains 26-30 are selected examples of
		neuronal spiking data from a neuronal culture recorded on a micro electrode array
		(see \ref{Appendix-Data-Kerstin}).
		All spike trains are normalized by their total length.
		\label{Fig:stereotypes}}
\end{figure}

To this aim, we use a test set composed of both artificial and real spike trains
(Fig. \ref{Fig:stereotypes}) to compare A-ISI-distance to ISI\hyp{}distance and
A-SPIKE-distance to SPIKE\hyp{}distance.
\textcolor{black}{We use two models to generate our samples.
For the spike trains with perfect periodicity we use a time varying steady rate
(fixed ISI) model.
For samples with more variability in spike timing we used a Poisson spiking model,
where the rate is fixed for a certain window at a time.
In some cases we add small jitter noise to both models.
The artificial spike trains 1-25 are designed to exhibit a variety of
stereotypical spiking behaviours including both single and multiple time scales.}
The experimental spike trains 26-30 consist of short recordings from neuronal
cultures on microelectrode arrays (see \ref{Appendix-Data-Kerstin} for details).
For the adaptive versions the threshold is estimated from the data (see Section
\ref{ss:MRTS_thr}) for each pair separately.

\begin{figure*}
	\centering
	\includegraphics[width = 0.8\textwidth]{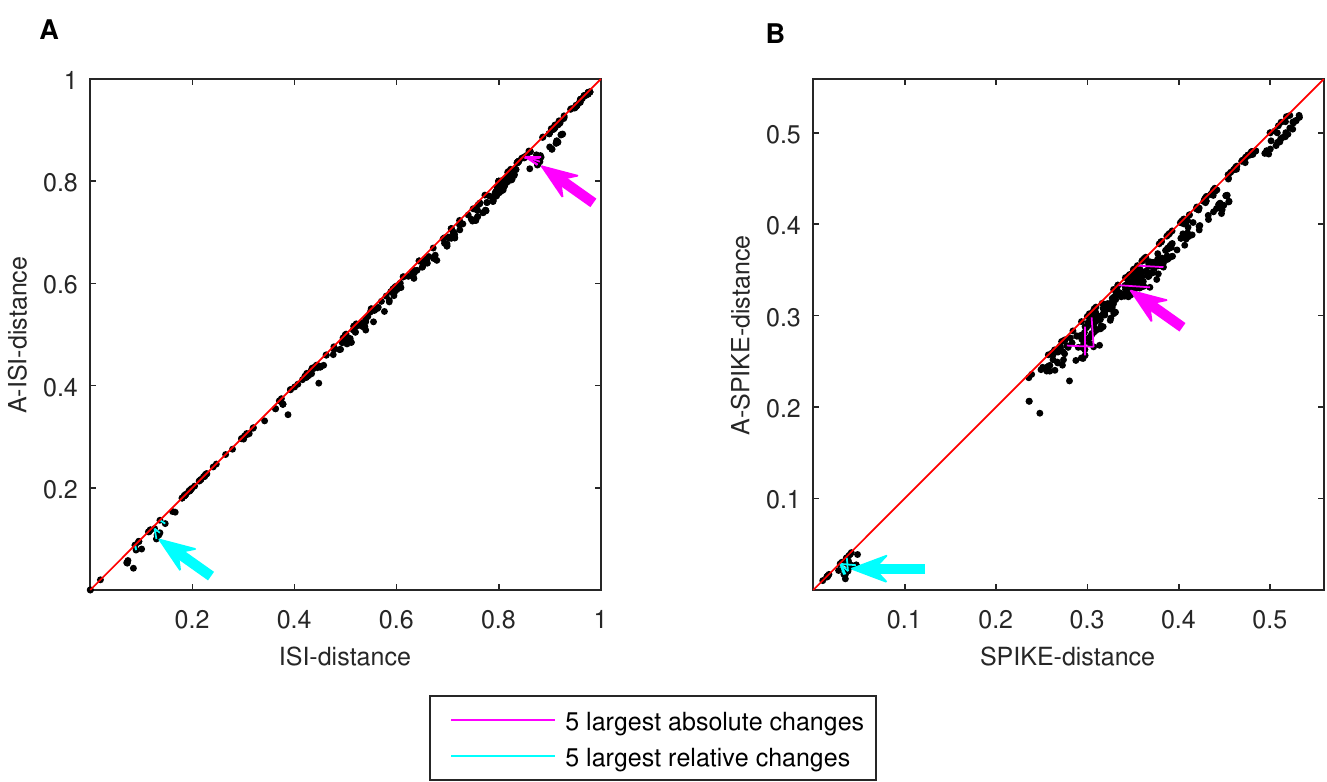}
	\caption[Figure]{
		\textcolor{black}{Scatter plots showing the A-ISI\hyp{}distance between all pairs of spike
		trains versus the original ISI-distance (A) and the A-SPIKE-distance versus the
		SPIKE-distance (B).
		The diagonal line marks where the measures would show equal distance.
		The pairs were sorted according to rising order of the original distances-
		Thus, if the order changed for adaptive extension, there is a negative slope in a
		line connecting all the pairs.
		For each line with a negative slope we calculated its length using the Euclidean 
		distance.
		The five largest absolute changes are indicated in magenta, the five largest
		relative changes in cyan.		
		In addition, a magenta (cyan) arrow points to the very largest absolute (relative) change.
		Overall, while the changes are seemingly small on an absolute scale, the relative changes
		can be very significant.
		\label{Fig:CombinedFig67}}}
\end{figure*}

In the analysis every spike train is paired with all the others.
Because for both the A-ISI-distance and the
A-SPIKE-distance the MRTS $\mathcal{T}$ can only reduce
but never increase the dissimilarity value, all pairs are found in the lower
half of the scatter plot (Fig. \ref{Fig:CombinedFig67}).
Furthermore, all values between pairs of spike trains are close to or on the diagonal,
which means that both versions attain very similar values or even the same value.
The differences between the two SPIKE\hyp{}distances are slightly more pronounced than
the differences between the two ISI\hyp{}distances.
Such seemingly small differences can still be of high significance since in a
typical experimental setup it is rarely the absolute value of similarity that matters
but rather the relative order of similarity for different conditions.
Moreover, in real data the range of similarity values obtained is usually quite small
which further increases the relative importance of small changes in similarity.

\begin{figure*}
	\centering
	\includegraphics[width = \textwidth]{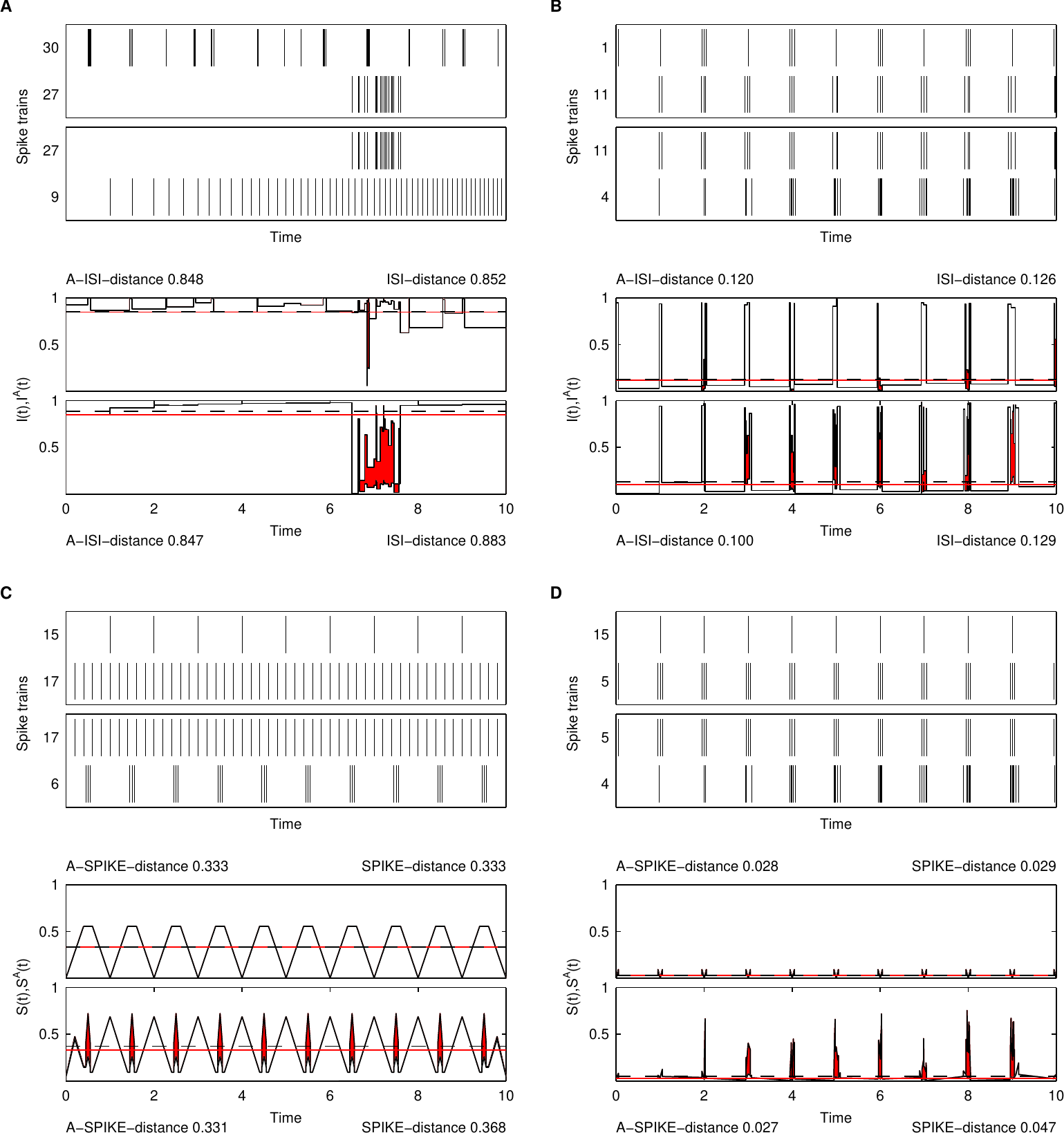}
	\caption[Figure]{
		\textcolor{black}{Largest absolute and largest relative change between ISI-distance and
		A-ISI\hyp{}distance (A,B) as well as SPIKE distance and A-SPIKE-distance (C,D) for the spike
		train set shown in Fig. \ref{Fig:stereotypes}.
		In both cases the two measures show a different order of similarity.
		The original distances increase from the first to the second pair, while the
		adaptive extensions decrease.	
		The first pairs attains distance values in between the second pairs, which results in a
		different order of similarity between the measures (see Fig. \ref{Fig:CombinedFig67}).	
		Upper subplots: The spike train pairs with the largest changes.
		Lower subplots: Respective original distance vs. adaptive version profiles with the
		difference between the two profiles emphasized.
		The distance values for the first (second) pair are shown on top (at the bottom).
		They are also marked by a dashed line for the original distance and by a solid line for
		the adaptive distance.
		\label{Fig:CombinedFig89}}}
\end{figure*}

For one spike train at a time we then look at all its pairings and sort the results
in ascending order according to the original versions.
The results from the adaptive versions are arranged in the same order.
If the order of the spike train pairs does not match, there is a clear difference
in the way the two measures consider spike train similarity.

We now investigate in more detail not only the largest absolute, but also the largest
relative changes observed in Fig. \ref{Fig:CombinedFig89}.
First, the largest absolute changes are identified by calculating the Euclidean distances
between the results for the two spike train pairs.
They typically take place for pairs of spike trains with large distances.
Next, since deviations from near perfect synchrony are more prominent and easier to
detect than differences between various levels of high dissimilarity, we also look
at relative changes.
These can be found by dividing each distance by its corresponding ISI- and A-ISI-distance
or SPIKE- and A-SPIKE-distance average.
For both distances they mostly occur for pairs of very similar spike trains.

For the A-ISI-distance, the spike train pairs showing the largest absolute change
compared to the ISI-distance can be seen in Fig. \ref{Fig:CombinedFig89}A.
The two measures show a different order of similarity; while the ISI\hyp{}distance
increases, the A-ISI-distance decreases from the first to the second pair. 
Spike trains 27 and 30 in the first pair are seen very similarly (deviation $<1\%$)
by both measures.
However, when spike train 27 is paired with spike train 9, the ISI\hyp{}distance
considers the local time scale only and thus has unreasonably high demands on the
spikes of the burst in spike train 27 which leads to large fluctuations in similarity.
For the A-ISI-distance on the other hand the burst matches another event
with relatively high rate and treats this event as a coinciding burst.
Outside of the burst the two measures agree that the spike trains are very dissimilar.

Fig. \ref{Fig:CombinedFig89}B depicts the two spike train pairs exhibiting the largest
relative change between the two ISI-distances.
While the first pair is seen as relatively similarly (deviation $<5\%$), the main difference
is found for the second spike train pair.
Here the ISI-distance looks at the detailed structure and judges the interspike intervals
within the bursts as very dissimilar, whereas the A-ISI-distance sees simply
matching bursts and attains a considerably lower distance value than the ISI-distance
(0.100 vs 0.129).

For the A-SPIKE-distance, the pairs of spike trains showing the largest absolute change
compared to the SPIKE-distance are depicted in Fig. \ref{Fig:CombinedFig89}C.
As there are no bursts in either of the two spike trains, both measures attain exactly
the same value for the first spike train pair.
This is a very good example for the specificity of the generalized version.
On the other hand, the original distance considers the second spike train pair (periodic
spiking versus periodic bursts) as much more dissimilar (increase $>10\%$).
In contrast to the SPIKE\hyp{}distance, it rightly considers the spike time differences
in the middle of two bursts as larger than the differences in the middle of the burst.

Finally, the largest relative change between the two SPIKE-distances is shown in
Fig. \ref{Fig:CombinedFig89}D.
Again, there is not much difference between the two distances for the first spike train pair.
However, the SPIKE\hyp{}distance considers the second spike train pair much more dissimilar
($>62\%$ higher) due to the large relative deviations in spike timing within their
coinciding bursts.
In contrast, the A-SPIKE-distance puts much less weight on the differences within bursts,
but still reacts to the spikes outside of the bursts.
This is an example of the sensitivity of A-SPIKE-distance.

All these results show that the effect of both generalized versions is strongest in
situations with multiple time scales in the spike trains.
A prominent example are bursts embedded in long silent periods.
In this case the long ISIs (of the silent periods) strongly influence the global 
time scales such that deviations of synchrony on the smallest time scales (within
the bursts) are weighted less.

\subsubsection{\label{ss:Results-A-SPIKE-sync}Adaptive SPIKE-synchronization}

A-SPIKE-synchronization can not be meaningfully tested by using the spike train set
of Fig. \ref{Fig:stereotypes}.
The perfect periodicity in many spike trains makes analysis of the
A-ISI-distance and the A-SPIKE-distance simple, but causes very abrupt changes in
A-SPIKE-synchronization due to its binary nature.
The values can be computed, but the largest differences are not meaningful with
this data set, since many spike trains with bursts jump from zero to a large
value and there is no way of ordering different pairs having zeros in the
original measure.
Thus, we here use a qualitative approach together with insights from the analysis of
A-ISI-distance and A-SPIKE-distance.

As a side effect of being time scale adaptive, SPIKE\hyp{}synchronization demands
very high spike timing accuracy during fast firing.
This leads to situations such as the one shown in Fig. \ref{fig:A-SYNC_simple}
(Section \ref{ss:SPIKE-sync}).
For spike trains 3 and 4 the spikes in the first event are considered coincident,
but the doublet in between them in spike train 2 is not judged as coincident with
either of them.
In contrast, for A-SPIKE-synchronization the coincidence windows are adapted to
the distribution of all ISIs in the data set and the two sides of the coincidence
window are allowed to be of different length.
With this change each of the spikes in the doublet becomes coincident with
one of the spikes (the respective closest one) in spike trains 3 and 4.

\begin{figure}[tb]
	\includegraphics[width = 0.48\textwidth]{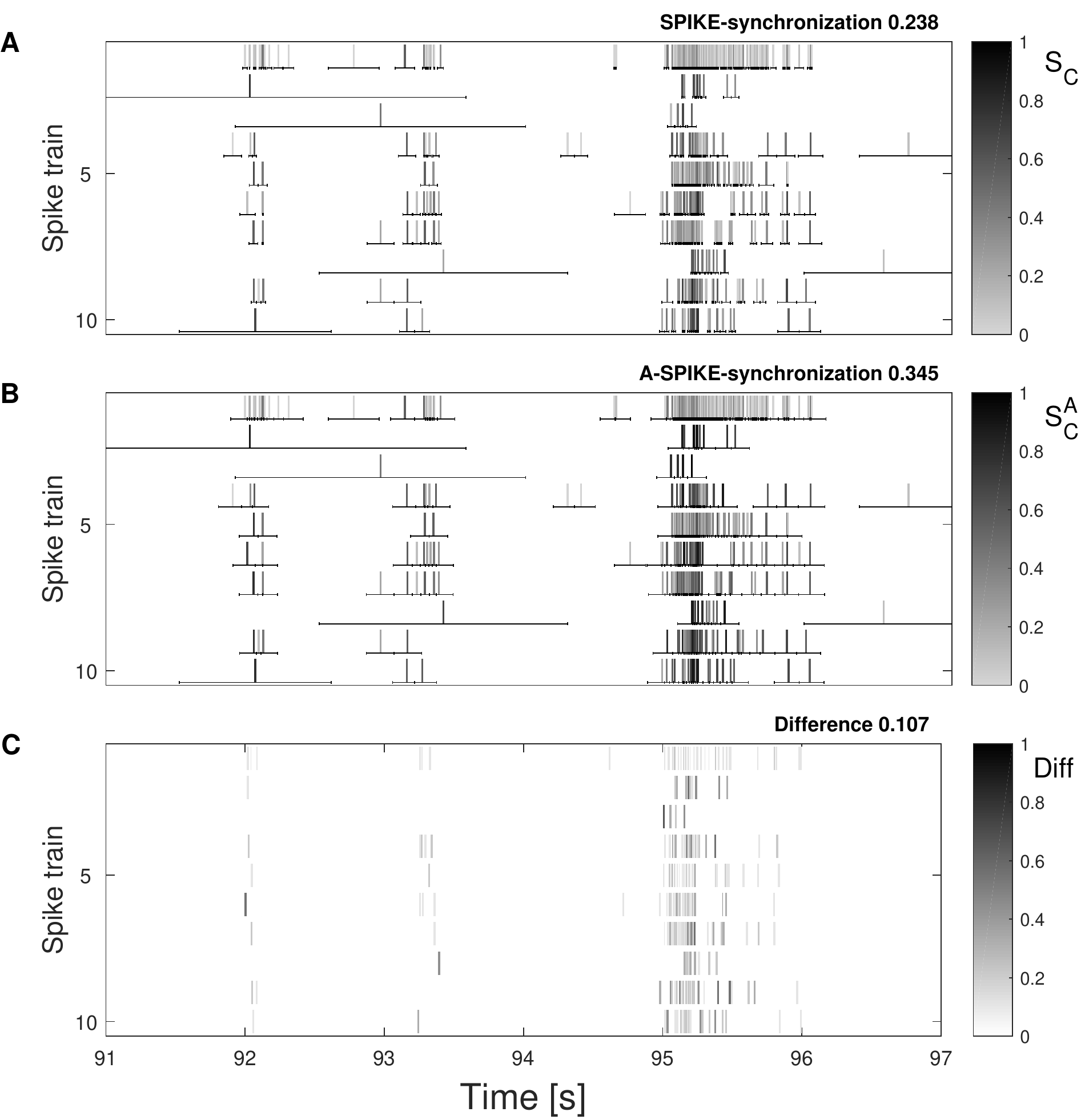}
	\caption{
		Real data example from MEA recordings (see \ref{Appendix-Data-Kerstin} for more details).
		Ten spike trains from the data set are plotted and their coincidence windows are
		drawn as obtained by SPIKE\hyp{}synchronization (A) and A-SPIKE-synchronization (B).
		The difference is plotted in C.
		Due to the adaptive coincidence windows, A-SPIKE-synchronization is able to match around
		45\% more spikes between bursting spike trains than SPIKE\hyp{}synchronization.
		As in Fig. \ref{fig:A-SYNC_simple}, the color scale is grey-black in A and B and
		white-black in C.
		\label{fig:A-SYNC_bursts}}
\end{figure}
\begin{figure*}[tb]
	\centering
	\includegraphics[width = 0.95\textwidth]{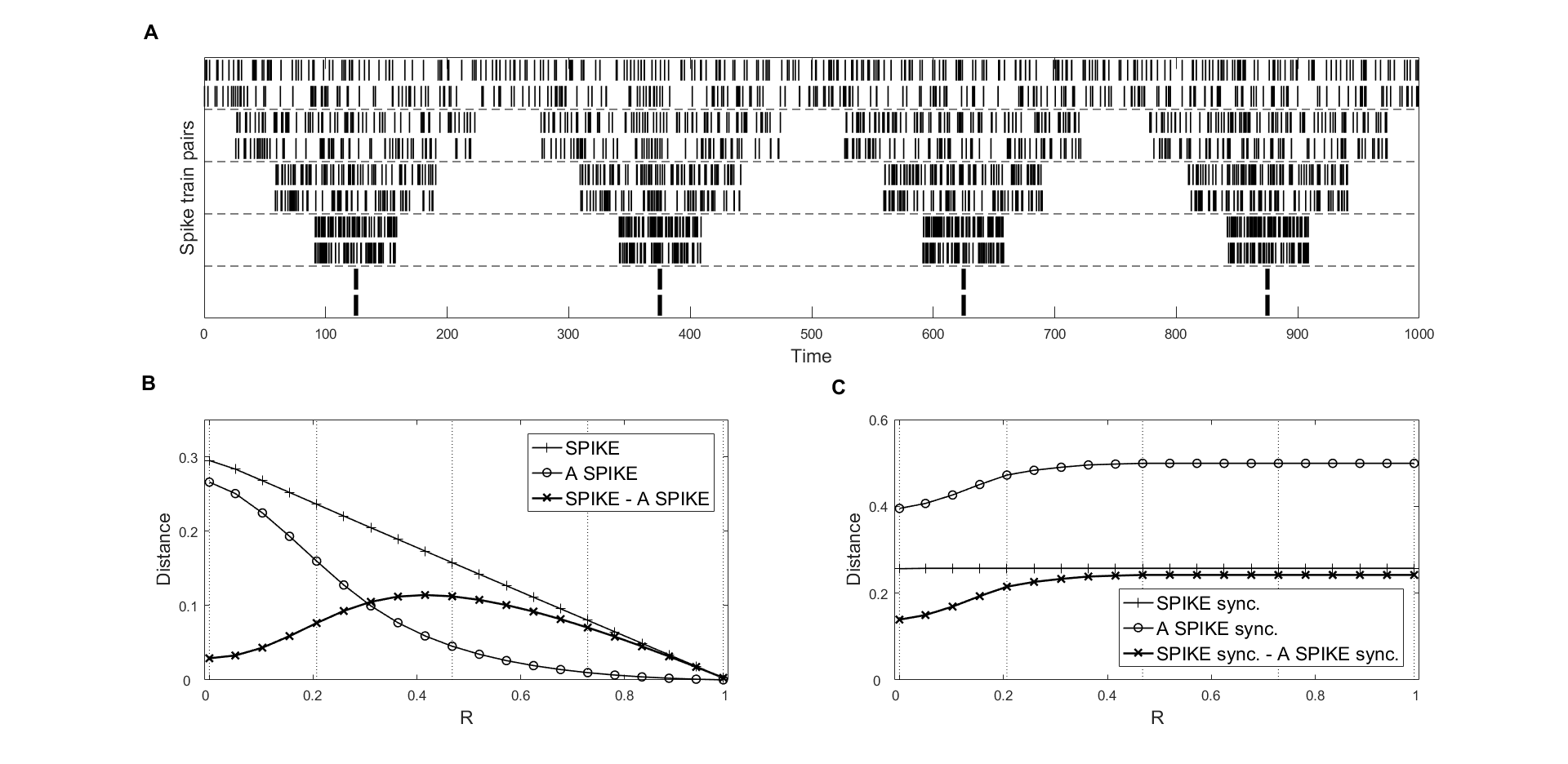}
	\caption[Figure]{
		 \textbf{\textcolor{black}{Effect of bursts on the adaptive versions evaluated by using the relative length $R$ of
		 interburst intervals.
		 The values are averages over 10 realizations.
		 (A) Five spike train pairs with increasing levels of burstiness for
		 one example realization.
		 (B) Effect of burstiness on the difference between A-SPIKE-distance and SPIKE-distance.
		 The graph for the ISI-distance looks very similar and is thus omitted.
		 (C) Equivalent results for A-SPIKE-synchronization.
		 The R-values of the examples in (A) are marked in (B) and (C) as dotted vertical lines.
		 }}
		\label{Fig:parameter_test}}
\end{figure*}
\begin{figure*}[tb]
	\centering
	\includegraphics[width = 0.95\textwidth]{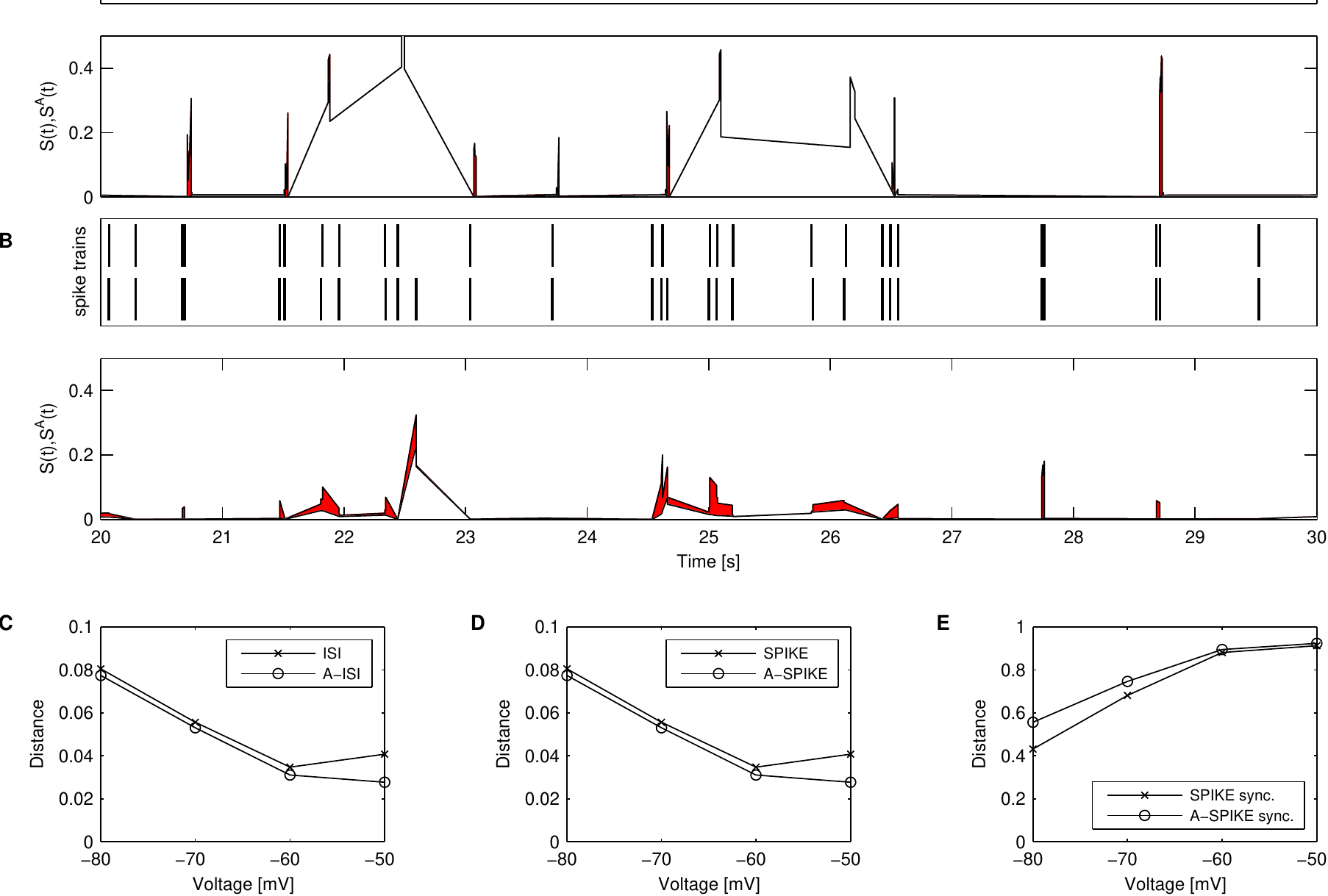}
	\caption[Figure]{
		\textbf{\textcolor{black}{Analysis of neuronal responses to multiple presentations of
		frozen noise for four different levels of the membrane potential.
		(A) Spike train responses (top) to two of the three noise presentations for a membrane
		potential of $-80mV$ and corresponding profiles for both A-SPIKE-distance and
		SPIKE-distance (bottom).
		The difference between the two profiles is marked in red.
		(B) Same as in (A) but for a membrane potential of $-50mV$.
		Results of the original and the adaptive measures for spike train sets of all three
		trials at four different voltage levels for the ISI-distance (C), the SPIKE-distance (D)
		and SPIKE-synchronization (E).}}
		\label{Fig:Real_Data}}
\end{figure*}
\begin{figure}[tb]
	\includegraphics[width = 0.46\textwidth]{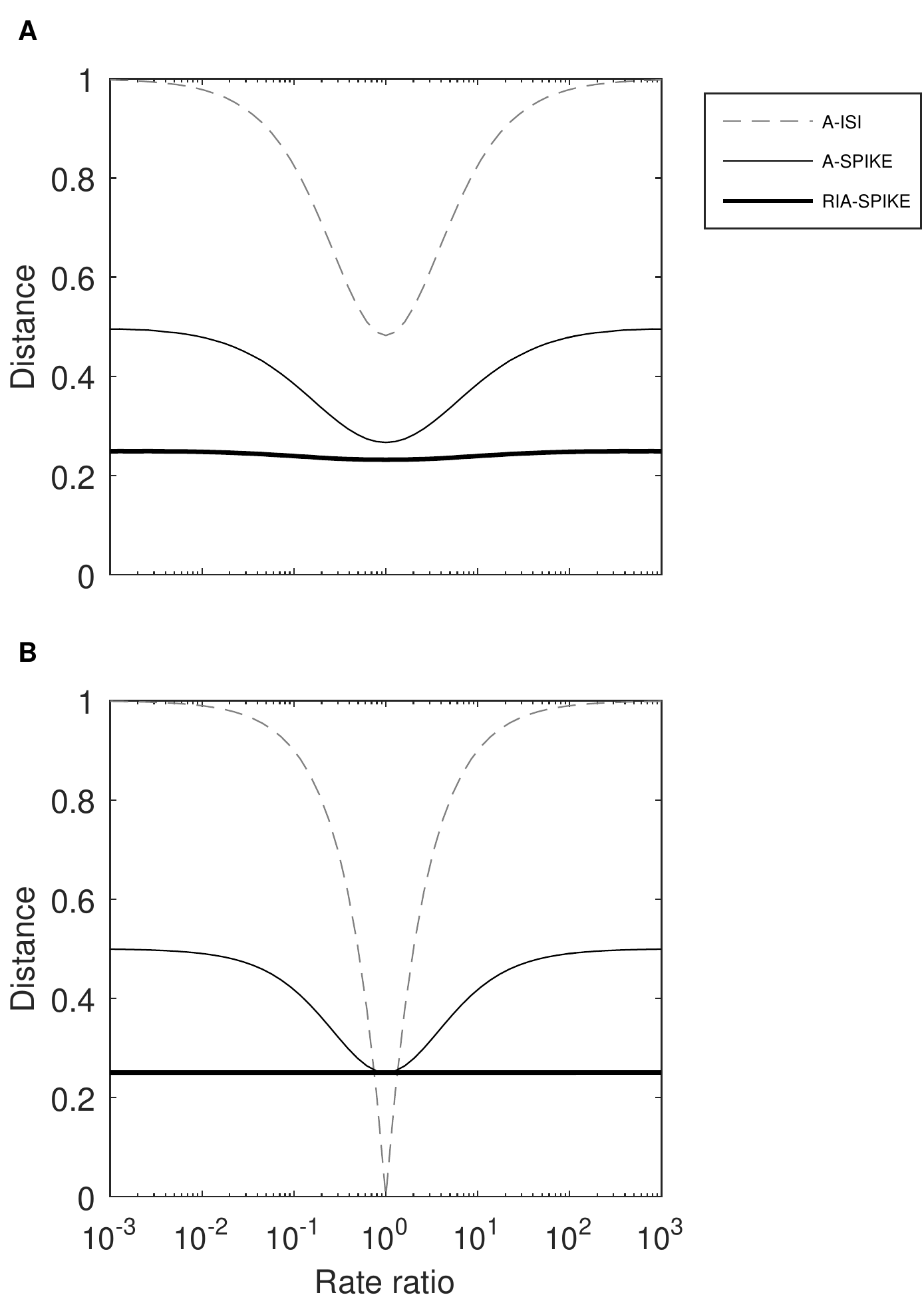}
	\caption[Figure]{
		Rate-independent RIA-SPIKE\hyp{}distance vs. A-SPIKE\hyp{}distance and A-ISI-
		distance for Poisson (A) and steady-rate spike trains (B).
		(A) Distances for two Poisson spike trains with varying rate ratios.
		The overall number of spikes in the two spike trains is kept constant.
		Each data point is an average over 100 trials.
		In contrast to the clearly rate-dependent A-ISI- and A-SPIKE\hyp{}distances, the
		rate-independent RIA-SPIKE\hyp{}distance exhibits an almost constant curve.
		(B) For the steady-rate spike train curves each data point is an average over 100 trials
		with random phase shifts between the two spike trains.
		In this case the line for the RIA-SPIKE\hyp{}distance is indeed perfectly constant.
		\label{Fig:rate_differences}}
\end{figure}

As a by-product of the adaptation, A-SPIKE-synchronization also increases the
coincidence window coverage within and at the edges of a burst and thus matches
as many spikes as possible.
For SPIKE-synchronization many of these spikes would be ignored due to the
unreasonably small coincidence windows.
This phenomenon occurs very often with real data.
An example containing two small and one large burst event is shown in Fig. \ref{fig:A-SYNC_bursts}.
In the first two events A-SPIKE-synchronization is able to detect a few additional
coincidences compared to SPIKE-synchronization.
The difference is much more pronounced for the third and largest event.
Here for SPIKE\hyp{}synchronization many potential matches are left out and this leads
to a rather low overall value of 0.238.
Instead, when A-SPIKE-synchronization is used, there are almost 45\% more matched spikes
within the burst and this strongly increases the overall synchronization value to 0.345.

Fig. \ref{fig:A-SYNC_bursts}C clearly shows that the additional spike matching of
A-SPIKE-synchronization only occurs in the high frequency events for which small
differences in the ISIs cause gaps between the coincidence windows of adjacent spikes.
Coincidences outside of these high frequency events are not affected.

\subsubsection{\label{ss:Results-Parameter-test} \textcolor{black}{Systematic evaluation of the influence of bursts}}

\textcolor{black}{Next we test how the effect of the automated threshold changes when spikes
are forming tighter bursts (Fig. \ref{Fig:parameter_test}).
To do this we first create two Poisson spike trains which are divided into four equally
long segments.
These segments are then increasingly compressed which prolongs the ISIs between them such
that the total length remains constant (Fig. \ref{Fig:parameter_test}A).
We use the relative length of the interburst intervals $R$ as a parameter and track the
difference between the adaptive and the original versions.
The results for the ISI-distance and the SPIKE-distance are very similar and we only show the latter.
From Fig. \ref{Fig:parameter_test}B we can see that the SPIKE-distance decreases almost
linearly with $R$ since the relative importance of the common silence in the interburst
intervals increases.
The adaptive version decreases sub-linearly with the largest absolute difference between
the two measures occurring around $R = 0.4$.
For higher $R$-values the reduction of the burst length overshadows the increases in similarity
at burst times and the difference increases up to a point and 
then starts to decrease.
The relative difference increases over the whole interval (data not shown, but can be
appreciated by observing the difference approaching the SPIKE-distance value towards $R = 1$).}

\textcolor{black}{While for the SPIKE-distance the interburst intervals have an effect
on the overall value, SPIKE-synchronization is sensitive to the matching of spikes only
and is based on one coincidence indicator value per spike.
Thus, the effect is increasing only until all possible spike pairs within the bursts are
matched.
For our example the increase saturates at $R = 0.4$ (Fig. \ref{Fig:parameter_test}C) at which
point all possible spike pairings (encompassing roughly half of the spikes) have been
identified.
This is in agreement with what we demand from a distance sensitive to bursting structure
for a systematic increase of the ratio between interbursts intervals and synchronous bursts.}

\subsubsection{\label{ss:Results-Real_Data} \textcolor{black}{Application to real data: Reliability of neurons}}

\textcolor{black}{In order to demonstrate the effects of the adaptive generalization 
in a more realistic scenario, we re-analyze data previously used to study the effect
of membrane potential resting state on neuronal reliability (\citet{Zeldenrust13},
see \ref{Appendix-Data-Fleur} for details on the recordings).
When in the original study frozen noise was injected into thalamocortical relay cells of rats,
it was found that the reliability of the cell response increases with depolarization
\citep{Zeldenrust13}.}

\textcolor{black}{Here we use both the original versions and the adaptive generalizations
of all three measures to assess the reliability of the responses from the two neurons
for which all four levels of membrane potential were recorded.
The adaptive versions use a threshold obtained from the data by pooling all spike
trains of each level and trial together.
In Fig. \ref{Fig:Real_Data} we show the results of the cell with the more prominent effect
but we get similar results for the other cell as well.
The cells analyzed were recorded three times for each holding membrane potential and reliability was assessed by trial to trial variations.
For the highest hyperpolarization (Fig. \ref{Fig:Real_Data}A) the original
SPIKE-distance yields spuriously high values for the local dissimilarity during
the bursts, since it only evaluates the local context.
Even when the A-SPIKE-distance takes the global context into account, both measures agree that there are
large dissimilarities in the spike trains.}

\textcolor{black}{For the most depolarized state (Fig. \ref{Fig:Real_Data}B) the
membrane potential is considerably closer to the action potential threshold.
The patterns are closely matching the burst positions of Fig. \ref{Fig:Real_Data}A,
but also additional events appear.
The neuron no longer responds in clearly distinguished bursts and it is
considerably more difficult to determine where a burst begins or ends.
Since the generalized version adapts to time scales found in all the spike trains,
it is able to distinguish when a burst-like pattern emerges and considers them as
more similar.}

\textcolor{black}{As can be seen in Figs. \ref{Fig:Real_Data}C and \ref{Fig:Real_Data}D, the original
versions, without adaptation and only using the local context, attain a higher
level of similarity for -60mV than for -50mV, which contradicts both the 
results in the original study and the results for SPIKE-synchronization (Fig. \ref{Fig:Real_Data}E).
Since the adaptive versions are able to make use of the global context of all the spike
trains, they attain results without this spurious dissimilarity and thus for higher
membrane potentials the similarity increases monotonously.}
	
\textcolor{black}{A-SPIKE-synchronization works slightly differently (Fig.
\ref{Fig:Real_Data}E).
Due to the tight bursts that cause excessively small coincidence windows, the largest
difference occurs for the hyperpolarized states.
However, both versions agree that the reliability as quantified by spike to spike matching
in the response patterns clearly show a monotonous increase over the baseline membrane
potential.
In summary, the results obtained by the A-SPIKE-distance and A-ISI-distance seem to be
appropriate and more in line with the original results.}

\section{\label{s:RI} Rate-independent extension}

Sometimes in neuroscience one is interested in the pure similarity of spike timing,
independent of any differences in spike rates.
Thus there is the need for a measure which can identify differences in spike timing
but is able to ignore any differences in rate between the spike trains.
Here we propose such a rate-independent extension for the A-SPIKE-distance.\footnote{The
A-ISI-distance is a measure of instantaneous rate difference and a rate-independent
measure of rate difference makes little sense.
A-SPIKE-synchronization is rate-dependent by definition, since it is calculated as the
average value of spike-based coincidence indicators (Eqs. \ref{eq:SPIKE-Coincidence}
and \ref{eq:Multi-SPIKE-Synchronization}).}

\subsection{\label{ss:RIA-SPIKE} RIA-SPIKE-distance}

In order to understand how rate-independence for A-SPIKE-distance is achieved,
we need to separate Eq. \ref{eq:A-SPIKE profile} (Section \ref{ss:SPIKE-dist})
for the pairwise A-SPIKE-distance profile into its three components.

The first two components are the mean of spike time dissimilarity and the normalization
to firing rate
\begin{equation} \label{eq:RI-SPIKE profile}
	S_{m,n}(t) = \frac{S_n(t)  + S_m(t) }{ 2 } \cdot \frac{1}{max\{\langle
	x_{\mathrm{ISI}}^{n,m}(t)\rangle, \mathcal{T} \}},
\end{equation}
where $S_n(t)$ and $S_m(t)$ are the weighted spike time differences for spike
trains $n$ and $m$ defined by Eq. \ref{eq:weighted distance}.
The third component is a weighting of the spike time dissimilarity according to the firing rate
difference that is applied to the first component
\begin{equation} \label{eq:Rate normalization}
	\frac{S_n(t) x_{\mathrm{ISI}}^m (t) +S_m(t) x_{\mathrm{ISI}}^n (t) }
	{\langle x_{\mathrm{ISI}}^{n,m} (t)\rangle}.
\end{equation}
The rate-independent adaptive SPIKE\hyp{}distance (RIA-SPIKE-distance) simply leaves out
this last weighting and can thus be written as
\begin{equation} \label{eq:RIA-SPIKE profile}
	S_{m,n}^{RIA}(t) = \frac{S_n(t)  + S_m(t) }{2\max(\langle x_{\mathrm{ISI}}^{n,m}
		(t)\rangle, \mathcal{T} )}.
\end{equation}
The RIA-SPIKE-distance shares all the properties of A-SPIKE-distance, but it only evaluates
normalized spike timing differences, whereas the A-SPIKE-distance additionally uses differences
in rate to determine similarity.


\subsection{\label{ss:RIResults}Results}

In this Section we compare the RIA-SPIKE-distance to the regular A-SPIKE-distance regarding
their response to differences in rate.
First, in Fig. \ref{Fig:rate_differences}A we look at Poisson spike trains with different
rate ratios.
The regular A-SPIKE-distance exhibits a clear rate dependency obtaining its lowest value for
spike trains with identical rates and increasing for higher rate differences.
The RIA-SPIKE-distance on the other hand starts near $0.25$ and remains relatively constant
for all rate ratios.
These deviations from perfect rate-independence occur because of the irregularities of the
Poisson spike trains.
When we repeat the same analysis with steady rate instead of Poisson spike trains (Fig.
\ref{Fig:rate_differences}B), thereby removing the effects of the Poisson statistics,
the RIA-SPIKE\hyp{}distance exhibits indeed perfect rate-independence.

Regarding the original distances, in Fig. \ref{Fig:rate_differences}A they would show 
very similar behavior to the adaptive generalizations.
Only for rate ratios close to $1$ there would be a small increase due to coincident
burst-like events within the Poisson spike trains.
In Fig. \ref{Fig:rate_differences}B the curves would overlap perfectly since there
is only one time scale in steady rate spike trains (both results not shown).

\section{\label{s:Discussion}Discussion}

In this manuscript we introduce adaptive generalizations to the three existing measures
ISI\hyp{}distance, SPIKE\hyp{}distance and SPIKE\hyp{}synchronization as well as a
rate-independent extension to the generalized SPIKE\hyp{}distance.
These new measures address two distinct problems.

The adaptive generalizations allow to disregard spike time differences that are not
relevant on a more global scale.
By means of a specifically constructed library of both stereotypical and real 
data spike trains, we can show that both A-ISI-distance and A-SPIKE-distance
indeed yield improvements for pairs of spike trains containing different time scales
without exhibiting any unwanted side effects in other examples.
Thus the changes are both sensitive and specific.
Regarding the size of the changes, even if they are seemingly small on an absolute scale,
the relative changes can be very significant.
For our test set the largest relative change reaches 29\% for the A-ISI-distance and even
up to 62\% for the A-SPIKE-distance.
With a more qualitative approach we then show that A-SPIKE-synchronization fixes the
problem of SPIKE-synchronization which demands an unreasonably high accuracy
for spike doublets and coinciding bursts.
By introducing a global reference frame, it manages to match spikes more efficiently (for
our test data we found an increase of 45\%).

\textcolor{black}{In order to test the adaptive measures methodologically we tested them in a controlled environment where two Poisson spike trains were split into bursts using increasingly large interburst intervals.
We designed the adaptive extension to be sensitive to bursting structure, therefore for
increasing relative length of interburst intervals we expect a larger difference between
the original and the adaptive versions. 
We show that the relative difference indeed increases monotonously with an increase in the
ratio between interbursts interval and bursts.}

\textcolor{black}{The absolute difference obtains its maximal value when the differences ignored in the bursts
are large and the bursts are long enough in comparison to the total length of the recording.
When very similar spike trains are compared their relative difference becomes dominant and internal
structures of coinciding bursts become less relevant.}
	
\textcolor{black}{
Additionally, we apply the measures to a dataset previously analyzed for reliability and find
that the adaptive methods agree with the previous results better than the original versions.
The A-ISI-distance and the A-SPIKE-distance seem to yield more reasonable results than the ISI-distance and the SPIKE-distance.
On the other hand when the coincidence windows of the original version get spuriously small,
A-SPIKE-synchronization can match spikes much more efficiently.
The effect can be especially meaningful in applications in which leader-follower relationships
based on the temporal order of spikes are determined \citep{Kreuz2016}.
}

The rate-independent extension on the other hand focuses on spike time accuracy while
disregarding rate differences in the two spike trains.
The original SPIKE\hyp{}distance considers spike time differences but also has a feature
that takes into account the firing rate difference between the spike trains.
However, sometimes only the spike time accuracy is of interest and for that purpose
the RIA-SPIKE-distance disregards any deviations in firing rate.
We can show that the RIA-SPIKE-distance is approximately rate-independent for Poisson
spike trains (apart from minor statistical effects) and perfectly rate-independent for
strictly periodic spike trains.
With this final addition we have completed the picture, since we now have measures that
are sensitive to rate only (A-ISI-distance), to timing only (ARI-SPIKE-distance), and to
both at the same time (A-SPIKE-distance).

The adaptive generalizations are implemented for cases where we have prior knowledge
of the system or where we want to reduce the importance of very small details.
However, one has to be careful with this method.
If the threshold parameter that defines the minimum relevant time scale (MRTS) is chosen
too high, this can introduce spurious synchrony.
To facilitate the selection, we introduce a method for automatically extracting the
threshold from the spike train data.
This is done by using the second moment the ISI-distribution of the whole dataset,
thereby giving more weight to longer ISIs.

Here it is important to note that while this automated estimation of MRTS gives us a
threshold value for each dataset, one has to be very careful when comparing results
obtained with different threshold values.
Thus, one cannot use the adaptive version for two recordings from the same source
without using the same threshold for both recordings, even if the ISI-distributions
differ.
In such cases, the preferable option would be to combine the ISI-distributions before
calculating the threshold and to use the resulting value for both recordings.
However, this might not work in all cases.
For example, recordings before and during an epileptic seizure can have very
different ISI-distributions.
This means that a globally meaningful threshold can not be extracted due to a very
bi-modal distribution of all the ISIs from the whole recording.
The resulting threshold would be in between the two modes which would cause the adaptive
measures to basically consider one of the recordings as a long burst and the other as
an almost silent period.
Thus, in cases where a suitable threshold can not be found, it is preferable to just set 
it to zero and consider only local information.
This is equivalent to using the original versions.

Many time scale parametric measures like the Victor-Purpura and the van Rossum distance
use a parameter to define the time scale of the system.
The threshold set for the adaptive versions is philosophically different in the sense
that it does not define a single time scale, but sets a line below which the effects
of the smaller time scales are being toned down.
All different time scales are still considered at the same time, but weighted
differently depending on how they compare to the threshold.

\textcolor{black}{Other measures that deal with multiple time scales exist.
For example, Lyttle and Fellous have proposed a metric to specifically assess the similarity of
spike trains with bursts or common silent periods \citep{Lyttle11}.
While in the proposed adaptive measures the time scale parameter is
limiting full time scale independence of the original measures, in
many measures the time scale is a fixed value. 
With the method proposed by Lyttle and Fellous they can detect bursts 
as well as silent periods. 
However, this comes with a cost, since the method requires two time 
scale parameters and three additional  parameters; length of minimum 
silent period, length of burst ISI, minimum number of spikes in a 
burst, scaling factor to decide how important bursts are in comparison 
to single spikes, and another factor to decide between importance of burst and 
silent period detection. 
While the large array of options gives the experimenter a powerful tool and provides more control over the analysis, 
it also increases the complexity of the overall experiment. 
This may cause problems, in particular when the data has many dimensions.
Similarly, Rusu and Florian have introduced a new class of metrics \citep{Rusu14}. 
The max-metric and the modulus-metric are well suited for measuring distances between spike trains where information is encoded
in bursts but single spike accuracy within burst is not relevant.
The max-metric depends on the kernel chosen and a time scale parameter deciding its size.
The modulus-metric is parameter free like the ISI-distance, the SPIKE-distance, and the SPIKE-synchronization.
This is achieved by using a very simplified kernel. 
However, the results obtained with both methods are not normalized. 
Thus based on the dissimilarity value alone it is not possible to say anything about the similarity of the two spike trains, but only about the order of different pairs.}

\textcolor{black}{Another often used alternative to spike train distances are correlation measures
(see e.g. \citeauthor{Cutts14}, \citeyear{Cutts14}).
However, these measures traditionally require windowing or binning and this creates
the problem that their performance can depend crucially on the window length or
bin size and also on the starting points and the overlap of the windows which clearly
reduces the objectivity of the results.}

The results confirmed our initial expectation that the main differences between the
adaptive generalizations and the original measures is in their assessment of the
similarity of bursty data.
Since bursts are ubiquitous and have been identified as an important area of neuroscience
research (see e.g. \citeauthor{Izhikevich2003}, \citeyear{Izhikevich2003};
\citeauthor{Sherman2001}, \citeyear{ Sherman2001}), there is a strong need for
this kind of similarity measurement.
For the ISI\hyp{}distance, a method has been proposed for evaluating the similarity of
bursty data by identifying bursts and assigning spikes at the beginning of the
bursts \citep{Qu16}.
However, burst detection is a notoriously difficult problem for which rather complicated
methods have been developed (see for example \citeauthor{Kapucu12}, \citeyear{Kapucu12}).
Thus, a measure based on assigning spikes to bursts inherits the problems of burst
detection.
Another problem with the measure proposed in \cite{Qu16} is that it disregards differences
in spiking behavior within the bursts.
In contrast, our adaptive versions do not detect bursts at all, but automatically
adapt their behavior whenever there are burst-like features in the data.

All the measures presented here are symmetric and thus invariant to the order of
the spike trains.
Recently we have developed a complementary directional approach consisting of
two new measures called SPIKE-order and Spike Train Order \citep{Kreuz2016}.
This approach utilizes the adaptive coincidence detection of SPIKE\hyp{}synchronization
to first sort multiple spike trains from leader to follower and then to quantify
the consistency of the spatio-temporal propagation patterns.
A natural continuation of the work presented in this article would be to use the
adaptive measures for this new approach as well.

We would like to finish by pointing out that the implementations of both the original
and the extended measures are provided online in three separate free code packages called
SPIKY\footnote[1]{http://www.fi.isc.cnr.it/users/thomas.kreuz/Source-Code/SPIKY.html} \citep{Kreuz15}
(Matlab GUI), PySpike\footnote[2]{http://mariomulansky.github.io/PySpike/}
(Python) \citep{Mulansky16} and, most recently, cSPIKE
\footnote[3]{http://www.fi.isc.cnr.it/users/thomas.kreuz/Source-Code/cSPIKE.html}
(Matlab command line with MEX-files).

\section*{\label{Acknowledgements} Acknowledgements }
\footnotesize
E.S., I.M., and T.K. acknowledge support from the European Union's Horizon 2020 research
and innovation program under the Marie Sklodowska-Curie Grant Agreement No. 642563
'Complex Oscillatory Systems: Modeling and Analysis' (COSMOS).
M.M., N.B., and T.K. acknowledge support from the European Commission through the
Marie Curie Initial Training Network 'Neural Engineering Transformative Technologies'
(NETT), project No. 289146.
K.L. was supported by the 3DNeuroN project in the European Union's
Seventh Framework Programme, Future and Emerging Technologies [Grant Agreement
No. 296590] and by the Tekes funded Human Spare Part project.
We thank Emre Kapucu, Inkeri V\"{a}lkki and Jari Hyttinen from the Hyttinen lab, BioMediTech,
Tampere, Finland for the data from the MEA-cultures,and Wytse Wadman and Pascal Chameau
from Cellular and Systems Neurobiology, Swammerdam Institute for Life Sciences, University
of Amsterdam, Amsterdam, the Netherlands for the current-clamp data.
Finally, we thank Ralph G. Andrzejak for useful discussions and for carefully reading
the manuscript. 

\normalsize

\appendix

\section{\label{Appendix-Methods}Edge effect correction and treatment of special cases}

Here, we deal with some subtle details in the definitions of all three measures
A-ISI-distance, A-SPIKE-distance and A-SPIKE-synchronization.
First, in \ref{ss:ISI_SPIKE_edge_corrections} and \ref{ss:SYNC_edge_corrections}, we correct
the edge effect by providing definitions for the periods before the first and after the
last spike in a spike train (for which the interspike interval is not defined).
This is necessary to guarantee that all measures are well-defined for the whole recording interval.
Subsequently, in \ref{ss:ISI_SPIKE_special} and \ref{ss:SYNC_special}, we deal with the two
special cases of empty spike trains and spike trains with only one spike.
Even if some spike trains are empty or very sparse, all measures should still be defined in a
way which is consistent with the regular definitions.

\subsection{Edge effect correction for A-ISI- and A-SPIKE-distance}\label{ss:ISI_SPIKE_edge_corrections}

Since the A-ISI- and the A-SPIKE-distance are time-resolved and are based on ISIs defined
by Eq. \ref{eq:ISI}, there is ambiguity at the edges before the first spike and after the
last spike.
To resolve this ambiguity we need to add auxiliary spikes. 
For the beginning of the spike train, we assign an auxiliary spike at the maximum of
the distance between the start of the observation interval and the first spike, and
the first known ISI
\begin{equation} \label{eqn:auxiliary_start}
	t_{s_{aux}}^{(n)} = t_1^{(n)} - \max\{t_1^{(n)}-t_s, t_2^{(n)}-t_1^{(n)}\}.
\end{equation}
This definition assumes that the rate stays the same at both sides of the spike
unless the edge is too far away for this to be true, in which case the auxiliary
spike is assigned at the edge.
Analogously, the time of the auxiliary spike at the end is
\begin{equation} \label{eqn:auxiliary_end}
	t_{e_{aux}}^{(n)} = t_{M}^{(n)} + \max\{t_e-t_M^{(n)},
	t_{M}^{(n)}-t_{M-1}^{(n)}\}.
\end{equation}
If the first or last spike is at the edge, no edge correction is necessary at that end.
This defines the ISI which is then used not only for the ISI\hyp{}distance but also for the
A-SPIKE-distance and A-SPIKE-synchronization.

An auxiliary spike used for the edge effect correction is basically treated as
any other spike, for example they can be the nearest neighbor to a real spike.
But there is one exception: In order to avoid artificial synchrony at the edges
in the A-SPIKE-distance, they use the distance to the nearest neighbor from the
first/last real spike
\begin{equation}\label{eqn:aux_two_or_more}
	\Delta t_{s_{aux}}^\nind{n} = \Delta t_1^\nind{n} \qquad\text{and}\qquad \Delta
	t_{e_{aux}}^\nind{n} = \Delta t_{M_n}^\nind{n}.
\end{equation}

\subsection{Edge effect correction for A-SPIKE-synchronization}\label{ss:SYNC_edge_corrections}

For the A-SPIKE-synchronization profile we first apply the edge effect correction
described above and then calculate the coincidence windows following Eqs. \ref{eq:ISI_before}
and \ref{eq:ISI_after}.

For cases when there is a spike right at the edge, we use the one ISI that
exists for setting the coincidence window of the spike to
\begin{equation} \label{eqn:one_isi_edge}
	\tau_1^{(n)} = \frac{1}{2}x_{1F}^{(n)} \qquad\text{and}\qquad \tau_{M}^{(n)} =
	\frac{1}{2}x_{MP}^{(n)}.
\end{equation}
We also determine that an auxiliary spike can under no circumstance be part of a
coincidence nor can it have a coincidence counter.
Finally, an auxiliary spike does not count as a spike in the normalization.

\subsection{Special cases for A-ISI- and A-SPIKE-distance}\label{ss:ISI_SPIKE_special}

Empty spike trains and spike trains with only one spike do not provide the
ingredients needed to apply Eq. \ref{eqn:auxiliary_start} and
\ref{eqn:auxiliary_end}.

In order to define the ISI of an empty spike train without any spikes, we assign
auxiliary spikes to its edges, the beginning and the end of the recording interval.
This is the only interval for which we can guarantee that there were no spikes.

However, while we can now use Eq. \ref{eq:ISI}, Eq. \ref{eqn:aux_two_or_more}
for the distance to the nearest neighbour of the auxiliary spikes is still ill-defined,
since there are no real spikes.
In this case a value is assigned exactly as in Eq. \ref{eq:Closest spike} and
the nearest neighbor can either be a real or another auxiliary spike.
A very reasonable implication of this definition is that two empty spike trains will be
considered equal by both measures.

Similarly, it is not possible to assess the rate at either side of a single spike.
The most reasonable auxiliary spike location is again at the edge of the recording.
Thus for both cases, the auxiliary spikes are assigned at the edges as
\begin{equation} \label{eqn:less_than_two_auxiliary}
	t_{s_{\textrm{aux}}}^{(n)} = t_s \qquad\text{and}\qquad t_{e_{\textrm{aux}}}^{(n)} = t_e
\end{equation}
and this completes the definitions for the A-ISI- and the A-SPIKE-distance.

\subsection{Special cases for A-SPIKE-synchronization}\label{ss:SYNC_special}

For A-SPIKE-synchronization the situation is slightly different, since it is not
continuous but only defined at the times of the spikes.
This means that by definition an empty spike train cannot have synchronous
spikes and thus has no value.
In case all spike trains are empty, we set A-SPIKE-synchronization to $S_C^A=1$,
i.e.\ empty spike trains are considered to be perfectly synchronous.
If a spike train contains only a single spike, we use half the spike train length to
define the coincidence window for the spike as
\begin{equation}
	\tau_{1}^{(n)} = \frac12 (t_e-t_s).
\end{equation}
These special cases complete the definition of A-SPIKE-synchronization.

\section{\label{Appendix-Data} Experimental recordings}

\subsection{\label{Appendix-Data-Kerstin} Microelectrode array recordings from mouse cortical
	cells}

The electrophysiological data analyzed in Sections \ref{ss:Results-A-ISI-SPIKE} and
\ref{ss:Results-A-SPIKE-sync} were recorded in the group of Prof. Jari Hyttinen at
Tampere University of Technology / BioMediTech, Tampere, Finland.
These recordings were performed prior to and independently from the design of this study.

Between 5,000 and 25,0000 commercially available primary mouse cortical cells (A15586,
Gibco, Thermo Fisher) were plated on five microelectrode arrays (MEAs; four
60MEA200/30iR and one 60HDMEA30/10iR, all purchased from Multi Channel Systems,
Reutlingen, Germany) following the protocol of Hales, Rolston, and Potter \citep{Hales2010}.
The dishes were coated with poly-L-lysine (Sigma-Aldrich, St. Louis, MO, USA)
and laminin (L2020-1MG, Sigma-Aldrich).
The medium for the MEA cultures was replaced three times a week.
All MEAs with cells were kept in an incubator (+37 $^{\circ}$C, 5\% CO2, 95\% air)
prior to and between recordings.
Data were recorded three times a week between the 4th and the 35th day in vitro.
Every recording lasted five minutes and was performed with 25 kHz sampling
rate.
Spike detection was carried out by setting an amplitude threshold at five times
the standard deviation of the signal-noise level and the spike time stamps were
stored with the Neuroshare Library for MATLAB (Multi Channel Systems).
We used two recordings for our examples and test sets.

The five real data spike trains used in the test set (spike trains 26 to 30 in
Fig. \ref{Fig:stereotypes}) were selected from these data by hand to represent
different time scales but chosen such that spike numbers were quite constant and
comparable to the artificial examples.	

\subsection{\label{Appendix-Data-Fleur} \textcolor{black}{Patch clamp recordings of rat thalamocortical relay cells}}

\textcolor{black}{The electrophysiological data analyzed in Section \ref{ss:Results-Real_Data}
were recorded at the Swammerdam Institute for Life Sciences, University of Amsterdam,
the Netherlands.
Again, these recordings were performed prior to and independently from the design
of this study.
The experiments carried out on brain slices from Wistar rats (Harlan, Netherlands;
postnatal days 12-16) were approved by the animal welfare committee of the
University of Amsterdam.}

\textcolor{black}{For details on the animals, slice preparation and electrophysiological recordings,
see \citet{Zeldenrust13}. 
In the current-clamp measurements the cell was injected with current that consisted of a
DC component with superimposed noise: a computer generated (MATLAB) time series of Gaussian
distributed random numbers of a length of 300 s, filtered by an exponential filter with
a time constant $\tau  = 10$ ms and a standard deviation of $\sigma  = 100$ pA.
A slow feedback system controlled the background DC current  to stabilize the membrane
voltage at one of the specified values (-80 mV, -70 mV, -60 mV or -50 mV) before the
actual recording started; after the start this DC current component was fixed.
The same frozen (= an exactly reproduced computer generated) noise train was injected into
the soma of the TCR neuron for every repetition of the experiment.
Signals were filtered at 5–10 kHz and sampled at 10–20 kHz.}

\textcolor{black}{The recordings consisted of trials from five different cells of which
only two included trials for all four levels of membrane potential.
The cells analyzed were recorded three times and reliability was assessed by
trial to trial variations.}


\end{document}